\documentclass[aps,prb,a4paper,twocolumn,superscriptaddress,longbibliography,nofootinbib]{revtex4-2}
\usepackage{amsfonts}
\usepackage{amsmath}
\usepackage{mathtools}
\usepackage{longtable}
\usepackage{amssymb}
\usepackage{graphicx}
\usepackage{bm}
\usepackage{color}
\usepackage{mathrsfs}
\usepackage[colorlinks,bookmarks=true,citecolor=blue,linkcolor=red,urlcolor=blue]{hyperref}
\usepackage{appendix}
\usepackage{float}
\usepackage{booktabs}
\usepackage[export]{adjustbox}

\newcommand{\RNum}[1]{\uppercase\expandafter{\romannumeral #1\relax}}

\begin{document}
\title{Absence of measurement-induced entanglement transition due to feedback-induced skin effect}
\author{Yu-Peng Wang}
\affiliation{Beijing National Laboratory for Condensed Matter Physics and Institute of Physics, Chinese Academy of Sciences, Beijing 100190, China}
\affiliation{University of Chinese Academy of Sciences, Beijing 100049, China}
\author{Chen Fang}
\email{cfang@iphy.ac.cn}
\affiliation{Beijing National Laboratory for Condensed Matter Physics and Institute of Physics, Chinese Academy of Sciences, Beijing 100190, China}
\affiliation{Songshan Lake Materials Laboratory, Dongguan, Guangdong 523808, China}
\affiliation{Kavli Institute for Theoretical Sciences, Chinese Academy of Sciences, Beijing 100190, China}
\author{Jie Ren}
\email{jieren@iphy.ac.cn}
\affiliation{Beijing National Laboratory for Condensed Matter Physics and Institute of Physics, Chinese Academy of Sciences, Beijing 100190, China}
\affiliation{University of Chinese Academy of Sciences, Beijing 100049, China}

\begin{abstract}
A quantum many-body system subject to unitary evolution and repeated local measurements with an increasing rate undergoes a measurement-induced entanglement transition from extensive (or subextensive) to area law entropy scaling.
We find that certain open boundary systems under ``generalized monitoring", consisting of ``projective monitoring" and conditional feedback, display an anomalous late-time particle concentration on the edge, reminiscent of the ``skin effect" in non-Hermitian systems.
Such feedback-induced skin effect will suppress the entanglement generation, rendering the system short-range entangled without measurement-induced entanglement transition.
While initially emerged in noninteracting models, such skin effect can also occur in chaotic interacting systems and Floquet quantum circuits subjected to random generalized measurements.
Since the dynamics of the skin effect do not require post selection, and can be observed at the particle number level, the phenomenon is experimentally relevant and accessible in noisy intermediate-scale quantum platforms, such as trapped ions.
\end{abstract}

\maketitle

\section{Introduction}
The competition between the measurement and unitary evolution produces a novel  \textit{measurement-induced entanglement transition} (MIET) \cite{PhysRevB.98.205136,PhysRevX.9.031009,PhysRevB.99.224307,PhysRevB.100.134306,Hastings2021,PRXQuantum.2.010352,puremeasure-2,Circuits-5,CFT-1,CFT-2,CFT-3}, where a random quantum circuit \cite{Circuits-1,Circuits-2,Circuits-3,Circuits-4,Circuits-5} interspersed by onsite measurements with an increasing rate goes from a volume-law regime to an area-law regime.
Similar entanglement transitions also appear in the context of monitored fermion chains \cite{FF-1,FF-2,FF-3,FF-4,FF-5,FF-6,longrange-1,longrange-2,Dirac-fermion}, monitored open systems \cite{opensystem-1,opensystem-2}, circuits with pure measurements \cite{puremeasure-1,puremeasure-2}, random tensor networks \cite{RTN-1,RTN-2,RTN-3,RTN-4}, and quantum error correction codes \cite{QECC-1,QECC-2,QECC-3,QECC-4}.
Measurements introduce intrinsic randomness to the otherwise deterministic dynamics, with each set of recorded measurement results corresponding to a specific trajectory \cite{trajectory}. 
Among various frameworks trying to explain the MIET \cite{stat-1,stat-2,stat-3,stat-4,Dirac-fermion,QECC-1,QECC-2,QECC-3,QECC-4,FF-2,PhysRevB.105.L241114,dual-1,dual-2,2207.12415}, one approach focuses on a specific trajectory, of which the evolution is described by a non-Hermitian Hamiltonian, where the entanglement transition in some systems coincides with the spontaneous PT symmetry breaking \cite{NH-1,NH-2,NH-3,NH-4,NH-5}.

One unique phenomenon in certain non-Hermitian open boundary systems is the \textit{non-Hermitian skin effect} \cite{NHSE-1,NHSE-2,NHSE-3,NHSE-4,NHSE-5,NHSE-6,NHSE-6,NHSE-7,NHSE-8}, where a finite portion of the eigenstates are spatially concentrated on the edges.
Dynamically, the skin effect implies that the late-time state from the quench dynamics will have particles concentrated near the edges.
The Pauli exclusion principle predicts a nearly tensor-product structure of the steady state which obeys the area-law entanglement scaling.
The skin effect in the non-Hermitian system has been shown to produce entanglement transition \cite{NH-4}.
On the other hand, some non-Hermitian systems, for example, the Hatano-Nelson model \cite{Hatano-Nelson}, display skin effect for any finite amount of non-Hermiticity, suggesting the absence of MIET and seemingly contradicting the putative universal entanglement transition in the monitored systems.
The non-Hermitian dynamics correspond to a particular trajectory, sometimes referred to as the no-click limit \cite{weakmeasure-2,trajectory}, which means all measurements return null results.
Implementing such dynamics will thus require post selection, where exponentially many experiments should be carried out before a desired trajectory is obtained.
In this work, however, we consider only the trajectory-averaged entanglement behavior of the whole ensemble of trajectories, which is more relevant to experimental implementation.
On the trajectory ensemble level, there will be no skin effect for conventional monitoring setups.
A natural question is whether the skin effect can appear in the whole ensemble of trajectories, and thus suppress the measurement-induced entanglement transition.

In this work, we propose a scheme to implement a special skin effect at the trajectory-averaged level, by introducing feedback operation into the monitoring dynamics.
Such feedback-induced skin effect (FISE) features a particle concentration on the edge (under open boundary conditions). 
The concentration of particle density can be characterized by a local order parameter called the classical entropy $S_\text{cl}$.
This quantity also imposes an upper bound on the bipartite entanglement entropy, i.e., $S_\text{cl} > 2 S_A$, where $S_A$ is the bipartite entanglement entropy for any subsystem $A$.
Numerical simulations of finite-size open boundary systems show a scale invariance of the steady-state classical entropy:
\begin{equation}\label{eq:intro-scaling}
	S_\mathrm{cl}(\gamma, L) = L\cdot G(\gamma L^{1/\nu}),\quad
	G(x \rightarrow \infty) \sim c/x^\nu,
\end{equation}
where $\gamma$ is the measurement rate and $L$ is the system size.
The scaling of Eq.~(\ref{eq:intro-scaling}) implies the saturation of the entanglement entropy in the thermodynamics limit at arbitrarily small measurement rates $\gamma$.
That is, for vanishing small $\gamma$, the scaling form in Eq.~(\ref{eq:intro-scaling}) implies $S_A < c/\gamma^\nu$, which is a large but constant value (does not scale with system size) and therefore the system is in the area-law entangled phase.
Remarkably, Eq.~(\ref{eq:intro-scaling}) persists even when the interaction is turned on, therefore eliminating (sub)extensive entangled phases believed to exist in the weakly-monitored regime.

Besides the continuous monitoring, FISE also appears in the randomly measured Floquet circuits, where the unitary evolution has only discrete spatial-temporal translation symmetry, and the monitoring process is the conventional projective measurement followed by a feedback operation.
Such quantum circuit models provide a novel type of dynamical skin effect, which is essentially a many-body phenomenon not described by any Hamiltonian.
The quantum circuit model is also experimentally relevant to the noisy intermediate-scale quantum devices.
As we will show in this work, starting from a  ``domain-wall" initial state, it takes a constant time (i.e., does not grow with system size) for the system to reach the steady state, and the skin effect is apparent even for a moderately small system size.
We propose to implement the feedback-induced skin effect on the trapped-ion platforms.

The structure of this paper is as follows.
In Sec.~\ref{monitoring-dynamics}, we analyze steady states of projective monitoring evolutions and motivate the introduction of feedback to achieve skin effect.
In Sec.~\ref{feedback-skin}, we investigate the phenomenologies of the feedback-induced skin effect and explore its influence on the measurement-induced entanglement transition.
In Sec.~\ref{floquet}, we extend the study of the feedback-induced skin effect to quantum circuit models with discrete spatial and temporal translation symmetry under random generalized measurements. 
Finally, we conclude our work in Sec.~\ref{conclusion}.

\section{Monitoring dynamics} 
\label{monitoring-dynamics}
\subsection{Projective monitoring} 

\begin{figure}
	\centering
	\includegraphics[width=0.85\linewidth]{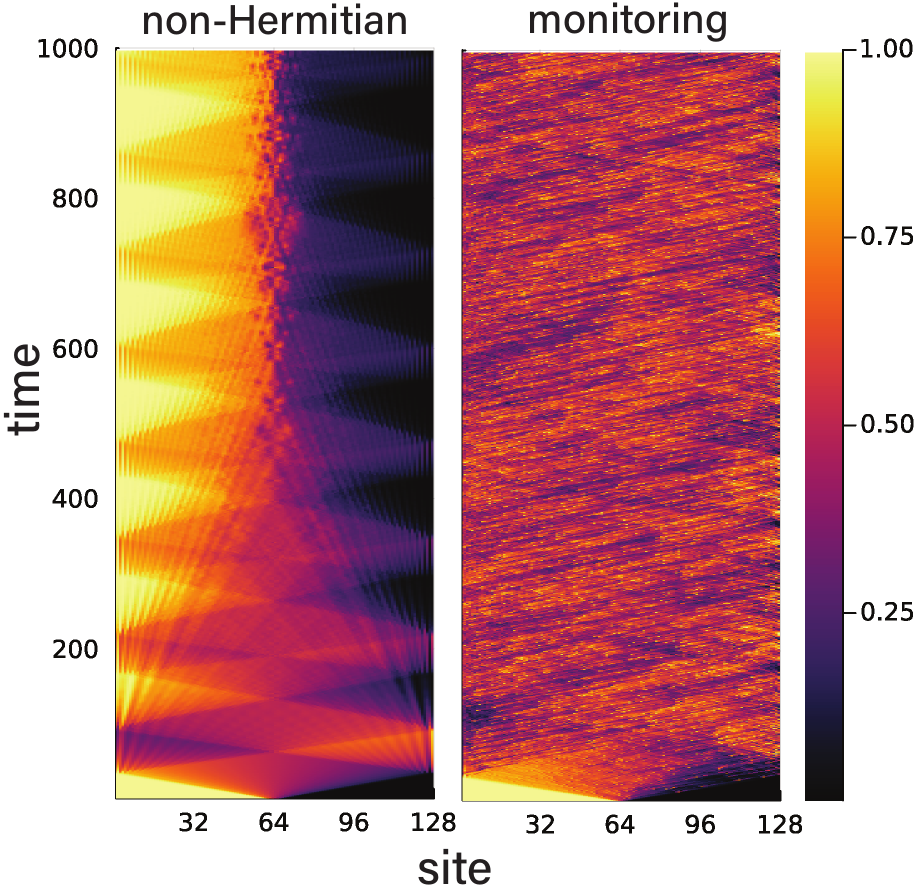}
	\caption{Comparison of non-Hermitian and monitoring dynamics starting from the initial state (\ref{eq:DW-state}) and under open boundary conditions. Left panel: Density evolution under non-Hermitian Hamiltonian with $\gamma=0.1$, where the particles gradually concentrate on the left. Right panel: Density evolution under projective monitoring with $\gamma=0.1$ (for a single set of measurement results). Although monitoring creates random quantum jump processes and thus introduces density fluctuation, the density distribution on a large scale is homogeneous.}
	\label{fig:compare}
\end{figure}

A standard quantum measurement process can be represented by a set of projectors ${P_m}$, each corresponding to a different measurement outcome. 
Measuring a state $|\psi\rangle$ yields a probability distribution over the possible outcomes, given by the Born rule: $p_m = \langle \psi| P_m |\psi\rangle$. 
The state then collapses to $P_m|\psi\rangle/\sqrt{p_m}$.
When the unitary evolution is governed by a Hamiltonian, we can consider a coarse-grained version of the dynamics known as \textit{projective monitoring}. 
This approach is formulated using the stochastic Schr\"{o}dinger equation (SSE) \cite{weakmeasure,weakmeasure-2,weakmeasure-3}:
\begin{equation}\label{eq:SSE}
\begin{aligned}
	d|\psi\rangle =&\ -i \left[H - \frac{i\gamma}{2} \sum_m (P_m - \langle \psi|P_m|\psi\rangle)\right] |\psi\rangle dt \\
    &\ + \sum_m \left[\frac{P_m |\psi\rangle}{\Vert P_m |\psi\rangle\Vert}-|\psi\rangle\right] dW_m,
\end{aligned}
\end{equation}
where each $dW_m$ is an independent Poisson random variable taking the values 0 or 1. 
In a small time interval $\Delta t$, the probability of observing $dW_m = 1$ (i.e., the probe registers a quantum jump) is proportional to $\gamma\Delta t$. 
If $dW_m=0$ for all $m$ (the no-click limit), the evolution is described by an effective non-Hermitian Hamiltonian 
\begin{equation}\label{eq:non-Hermitian}
	H_\text{eff} = H -i \frac{\gamma}{2} \sum_m P_m.
\end{equation}
In this section, we first consider a spinless fermion chain described by a nearest-neighbor hopping Hamiltonian: 
\begin{equation}
	H = \sum_i (c_i^\dagger c_{i+1} + c_{i+1}^\dagger c_i).
\end{equation}
The observable of interest in this system is the occupation number of local quasimodes created by two-site projectors:
\begin{equation}
	P_m = d_m^\dagger d_m, \quad
	d_m^\dagger = \frac{1}{\sqrt 2}(c_m^\dagger-i c_{m+1}^\dagger),
\end{equation}
where $m = 1,2,\dots,L-1$.
Note that the quasimode $d^\dagger_m$ is a right-moving wave packet, which is apparent in the momentum space: $d^\dagger_m = \sum_k f_m(k) c_k^\dagger$, where $|f_m(k)|^2= \frac{1}{4}(1+\sin k)$ peaks at $k=\pi/2$.
The effective non-Hermitian Hamiltonian (\ref{eq:non-Hermitian}) for the monitored system is given by 
\begin{equation}\label{eq:Hatano-Nelson}
\begin{aligned}
	H_\mathrm{eff} = &\ \sum_i \left[\left(1+\frac{\gamma}{4}\right) c_i^\dagger c_{i+1} + \left(1-\frac{\gamma}{4}\right) c_{i+1}^\dagger c_i \right] \\
	&\ -i\frac{\gamma}{4}\sum_i (n_i + n_{i+1}).
\end{aligned}
\end{equation}
This Hamiltonian is known as the Hatano-Nelson model \cite{Hatano-Nelson} and displays the non-Hermitian skin effect.
Consider the evolution of the ``domain-wall" state:
\begin{equation}\label{eq:DW-state}
	|\psi_0\rangle \equiv |1\cdots10\cdots0\rangle.
\end{equation}
In the no-click limit and under open boundary conditions, the effective non-Hermitian Hamiltonian (\ref{eq:Hatano-Nelson}) results in a late-time particle accumulation on the left edge (as displayed on the left panel of Fig.~\ref{fig:compare}).
On the contrary, in the projective monitoring case, the detected right-moving quasimodes will balance out the momentum distribution, leaving a steady state of homogeneity (as displayed on the right panel of Fig.~\ref{fig:compare}). 

Besides SSE, the trajectory-averaged dynamics can also be formulated as the Lindblad master equation \cite{Lindblad-1,Lindblad-2,openquantumsystem}:
\begin{equation}\label{eq:Lindblad}
	\frac{d}{dt}\rho = -i[H,\rho] -\frac{\gamma}{2}\sum_m \{L_m^\dagger L_m, \rho\} + \gamma\sum_m L_m\rho L_m^\dagger,
\end{equation}
where $\overline{|\psi\rangle\langle\psi|}$ is the density matrix averaged over all trajectories, and $L_m = P_m$ is referred to as the \textit{jump operator}.
A state $\rho_{\mathrm{NESS}}$ is a nonequilibrium steady state (NESS) if $\partial_t \rho_{\mathrm{NESS}} = 0$.
Note that for the projective monitored dynamics, all jump operators $\{P_i\}$ are Hermitian.
The maximally mixed state within a given particle-number sector $\rho = (\operatorname{Tr} \mathbb I_\nu)^{-1} {\mathbb I}_{\nu}$ (the subscript indicates the subspace spanned by the states of filling number $\nu$) is automatically a steady state since
\begin{equation}
\begin{aligned}
	\frac{d}{dt}{\mathbb I}_{\nu} &= -\frac{\gamma}{2}\sum_m \{P_m, {\mathbb I}_{\nu}\} + \gamma\sum_m P_m {\mathbb I}_{\nu} P_m^\dagger \\
	&= -\gamma \sum_m \left.P_m\right|_\nu + \gamma \sum_m \left.P_m\right|_\nu = 0.
\end{aligned}
\end{equation}
In general, there can be multiple steady states in each particle-number sector.
However, even for degenerate cases, the degeneracy can be lifted by adding boundary perturbation while preserving the bulk property of the system.\footnote{The uniqueness of steady state can be proved using the method in  Refs.~\cite{uniqueness-1,uniqueness-2} (see also a review in Ref.~\cite{uniqueness-3} and an application in Ref.~\cite{uniqueness-4}, where the system is under boundary driven \cite{PhysRevLett.123.230604}), which proved that a Lindblad equation has unique nonequilibrium steady state if and only if the set $$\mathcal O \equiv \{H, L_1, L_1^\dagger,L_2,L_2^\dagger,\cdots\}$$ generates (under multiplication and addition) the complete algebra on the Hilbert space. The general proof assumes no conserved quantity for the Lindblad equation. For the particle number conserving case, as we considered in the main text, we can focus on the Hilbert subspace $\mathcal H_\nu$ spanned by states will filling number $\nu$. The uniqueness condition then says if $\mathcal O|_{\mathcal H_\nu}$ generates the complete algebra on $\mathcal H_\nu$, then the steady state in $\mathcal H_\nu$ will be unique. We prove the uniqueness of the steady state $\rho_\mathrm{NESS}$ for the specific model considered in this work in Appendix~\ref{uniqueness}.}
Assuming nondegeneracy, the steady states of the projective monitoring evolutions should be the featureless maximally mixed states, and therefore we expect no skin-effect-like dynamics.

We also show in Appendix~\ref{apx:proj-monitor} that the projective monitored systems, under open boundary conditions, show conventional measurement-induced entanglement transition.
In this case, in the small-$\gamma$ regime, the particular trajectory in the no-click limit becomes a statistical outlier in terms of the entanglement entropy, which fails to represent the entanglement transition of the ensemble.
In the following, we will show that certain feedback will drive the rest of the trajectories to a similar skin-effect configuration, and therefore suppress the measurement-induced entanglement transition.

\subsection{Generalized monitoring} 
It is important to note that projective monitoring is an idealized representation of an actual measurement process.
In practice, detecting a quantum state requires a probe to interact with the system, which inevitably disturbs the measured states. The more general form of a quantum measurement is described by the \textit{positive operator-valued measure} (POVM) formalism \cite{nielsen_chuang_2010,wilde_2013}. 
A continuous version of the POVM called the generalized monitoring, is formulated as the SSE in Eq.(\ref{eq:SSE}), with the projector $P_m$ replaced by a general operator $L_m$ (see Appendix~\ref{generalized measurement} for a short review).
In this study, we focus on a particular form of SSE where 
\begin{equation}
	L_m = U_m P_m,
\end{equation}
corresponding to adding a unitary feedback operator $U_m$ to the projective monitoring process. 
The conditional feedback does not affect the effective non-Hermitian Hamiltonian $H_\mathrm{eff}$, but instead operates on those trajectories that deviated from the post-selected trajectory. 

\begin{figure*}
	\centering
	\includegraphics[width=0.95\linewidth]{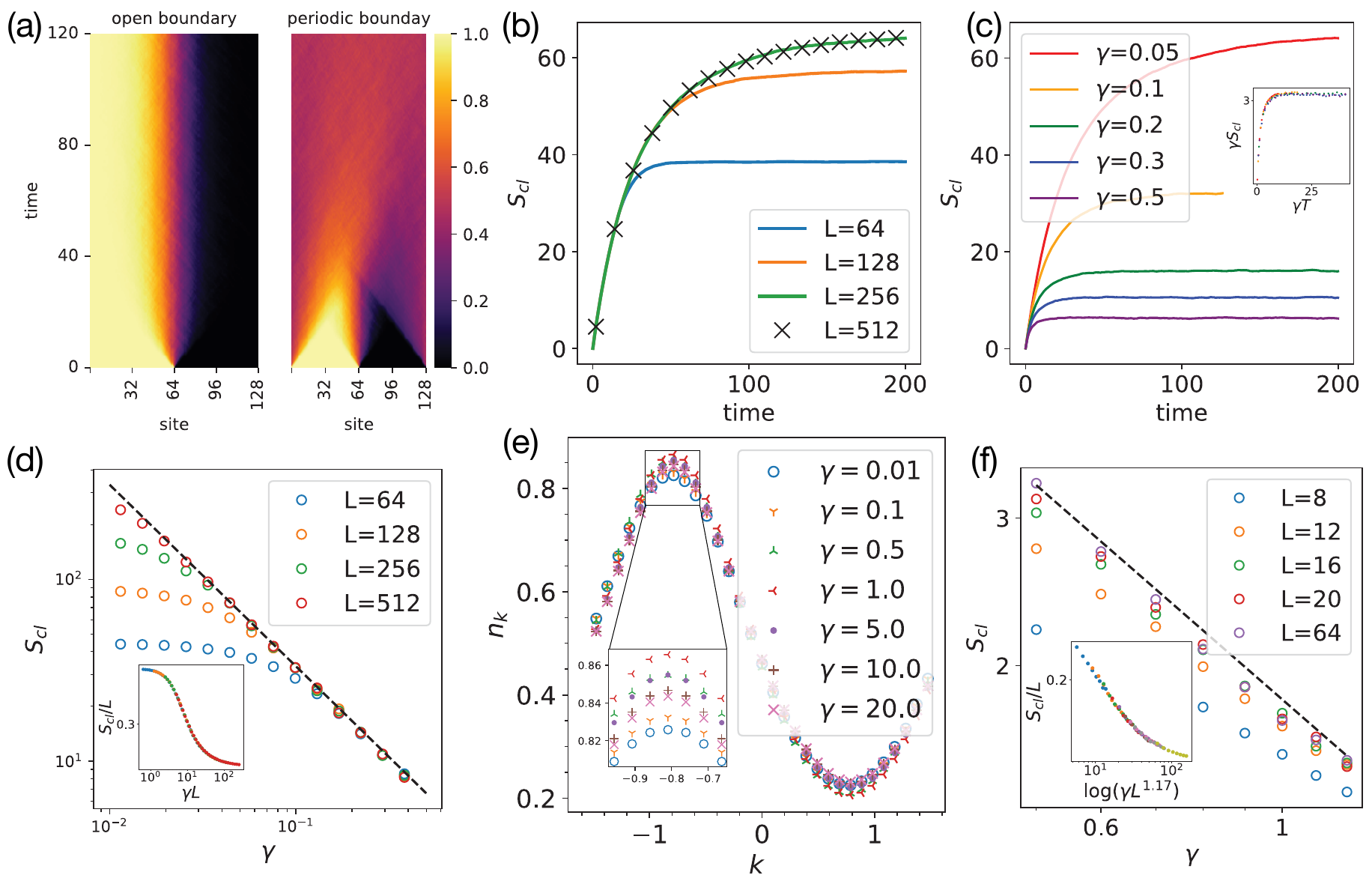}
	\caption{
	(a) Mean particle density evolutions of generalized monitored free fermions ($L=128$, $\gamma=0.1$) under open boundary conditions (left panel) and periodic boundary conditions (right panel). The feedback induces a strong boundary sensitivity: Under OBC the particle density evolves to be evenly distributed, while under OBC the steady state still features the domain structure.
	(b) Mean dynamics of $S_\text{cl}$ for $\gamma=0.05$ for different system sizes. The final saturation values of $S_\text{cl}$ depend on the system size $L$ when $L$ is no bigger than the domain wall length $L_\text{DW}$. While for sufficiently large size ($L \ge 256$ in this case), $S_\text{cl}(t)$ shows no system size dependence.
	(c) Mean dynamics of $S_\text{cl}$ for $L=512$ ($L > L_\text{DW}$ for all $\gamma$'s we choose). The data collapse indicates that the saturated evolution $S_\text{cl}(t;L>L_\text{DW})$ fits into the universal form Eq.~(\ref{eq:scaling-form-1}). 
	(d) Finite-size scaling of steady-state $S_\text{cl}$, which approach the line $S_\text{cl} = c/\gamma$ where $c \approx 3.3$. Inset: Data collapse indicates the scaling form $S_\text{cl} = L \cdot G(\gamma L)$.
	(e) Momentum distribution of the steady state, in the periodic boundary conditions.
	(f) Steady-state $S_\text{cl}$ for interacting monitored model, which approach the line $S_\text{cl} = \tilde c/\gamma^{\nu}$ where $\tilde c \approx 1.8$ and $\nu=0.85$. The data collapse indicates the universal scaling form $S_\text{cl} = L \cdot \tilde G(\gamma L^{1/\nu})$.}
	\label{Fig}
\end{figure*}

Consider a conditional unitary operator
\begin{equation}
	U_m = \exp(i \theta n_{m+1}),
\end{equation}
acting on sites $m+1$ whenever the probe detects a right-moving mode $d^\dagger_m$.
We temporarily fix $\theta=\pi$ so that the feedback $U_m$ converts the detected right-moving mode to a left-moving mode:
\begin{equation}
	\tilde{d}_m^\dagger\equiv U_m d_m^\dagger U_m^\dagger = \frac{1}{\sqrt 2}(c_m^\dagger+ic_{m+1}^\dagger).
\end{equation}
Consequently, the monitored dynamics with feedback always increase left-moving particles, resulting in the FISE where particles concentrate on the left boundary. 

As displayed on the left panel of Fig.~\ref{Fig}(a), starting from $|\psi_0\rangle$ with a sharp domain wall, the late-time dynamics still feature two static domains where $\langle n_i \rangle$ only takes the extreme value of $1$ or $0$.
The particle exchange happens only in the vicinity of the border, where the sharp domain wall gradually blurs to a fuzzy domain-wall region.
Comparing this with the dynamics with the periodic boundary conditions [right panel of Fig.~\ref{Fig}(a)], where particles quickly disperse into a homogeneous state, we see the FISE features an anomalous boundary sensitivity.

\section{Feedback-induced skin effect}
\label{feedback-skin}

\subsection{Dynamical scale invariance}

To quantify the dynamical skin effect induced by feedback, we introduce the classical entropy:
\begin{equation}
	S_\mathrm{cl}[\{n_i\}] \equiv - \sum_{i}\left[n_i \log n_i + (1- n_i)\log(1- n_i)\right].
\end{equation}
where only the nontrivial density (i.e., $n_i\neq 0,1$) contributes to $S_\mathrm{cl}$.

In Fig.~\ref{Fig}(b), we show the time evolution of the classical entropies $S_\mathrm{cl}[\{\overline{\langle n_i \rangle}\}]$ starting from $|\psi_0\rangle$ under open boundary conditions.
We see when the system size is small, $S_\text{cl}$ will saturate to a system-size-dependent value.
Whereas for large enough systems, the evolution of $S_\text{cl}(t)$ will be the same. 
These numerical simulations support the ``domain-wall blurring" picture, where the evolution of $S_\mathrm{cl}(t)$ follows a universal pattern when considering a system larger than the size of the domain-wall region (denoted as $L_\text{DW}$).

In Fig.~\ref{Fig}(c), we display $S_\text{cl}(t)$ for $L > L_\text{DW}$ systems.
We observed a scaling form
\begin{equation}\label{eq:scaling-form-1}
	S_\text{cl}(\gamma,t, L > L_\text{DW}) = \frac{1}{\gamma} F(\gamma t),
\end{equation}
where the system reaches the steady state with a characteristic relaxation time $t_\text{rlx} \sim \gamma^{-1}$, independent of the system size.

Furthermore, we observe that the steady-state $S_\text{cl}$ has a scale invariance, as shown in Fig.~\ref{Fig}(d):
\begin{equation}\label{eq:scaling-form-2}
	S_\mathrm{cl}(\gamma, t > t_\text{rlx}, L) = L \cdot G(\gamma L),
\end{equation}
which implies that $L_\text{DW} \sim \gamma^{-1}$.
In the late-time and thermodynamics limit, the asymptotic behaviors of the scaling functions are 
\begin{equation}
	F(x\rightarrow \infty) \sim c,\quad 
	G(x \rightarrow \infty) \sim \frac{c}{x},
\end{equation}
where $c$ is a numerical constant estimated to be $3.3$ in this specific case.
Therefore,
\begin{equation}\label{eq:scaling-form-3}
	S_\mathrm{cl}(\gamma, t > t_\text{rlx}, L > L_\text{DW}) = \frac{c}{\gamma} \approx \frac{3.3}{\gamma}.
\end{equation}
Notably, Eq.~(\ref{eq:scaling-form-2}) suggests a scale invariance for the steady-state density profile $n(x)$ in the continuum limit.\footnote{In contrast, when $\gamma$ is sufficiently large ($\gamma \sim O(c)$), the lattice effect becomes prominent, and the scaling behavior breaks down.}
The finite value of $S_\text{cl}$ implies that the feedback-induced skin effect is present even for a vanishingly small measuring rate.

\subsection{Suppression of entanglement}
One of the consequences of FISE is the suppression of entanglement.
Specifically, we prove that $S_\mathrm{cl}$ imposes an upper bound for the (trajectory-averaged) steady-state entropy of the monitored dynamics.
Consider an arbitrary subsystem $A$ inside the monitored system.
The entanglement \textit{subadditivity} \cite{nielsen_chuang_2010,wilde_2013} leads to the inequality 
\begin{equation}
	S_A \le \sum_i S_i.
\end{equation}
Further, consider the reduced density matrix of a local spin at site $i$:
\begin{equation}
	\rho_i = \begin{bmatrix}
		n_i & o_i \\ o_i^* & 1-n_i
	\end{bmatrix},
\end{equation}
where $o_i$ is the off-diagonal element.
Mathematically, the eigenvalues of a positive $2\times 2$ matrix \textit{majorize} \cite{matrix} the diagonal elements, and thus have less entropy. 
That is,
\begin{equation}
\begin{aligned}
	S_i &= - \operatorname{Tr} [\rho_i\log\rho_i] \\
	&\le n_i \log n_i + (1- n_i)\log(1- n_i) \\
	&= S_\text{cl}[n_i].
\end{aligned}
\end{equation}
Therefore, the entanglement entropy of subregion $A$ is bounded by the classical entropy
\begin{equation}
	S_A \le \sum_{i \in A} S_\text{cl}[n_i] = S_\text{cl}[\{n_i|i\in A\}].
\end{equation}
Also, for a pure state, the entanglement entropy of $A$ is equal to that of its complement $\bar A$, i.e., $S_A=S_{\bar A}$.
The classical bound on entanglement entropy also requires
\begin{equation}
	S_A + S_{\bar A} \le S_\text{cl}[\{n_i\}],
\end{equation}
therefore $S_A \le S_\mathrm{cl}/2$.

In the context of entanglement transition in monitored systems, it is natural to consider the trajectory-averaged entanglement, denoted as $\overline{S_A}$, which is bounded by $\overline{S_\text{cl}[\{\langle n_i \rangle\}]}/2$.
Since the entropy function is convex, $\overline{S_A}$ will also be bounded by:
\begin{equation}
	\overline{S_A} \le \frac{1}{2}\overline{S_\text{cl}[\{\langle n_i \rangle\}]}
	\le \frac{1}{2} S_\text{cl}\left[\overline{\langle n_i\rangle}\right].
\end{equation}
The asymptotic behavior, therefore, predicts the area-law entanglement scaling $S \le c/2\gamma$ for arbitrary $L$, thus proving the area-law entanglement scaling in the $\gamma \rightarrow 0$ limit.

To complete the analysis, in Appendix~\ref{numerics-gen}, we explicitly calculate the trajectory-averaged entanglement entropy (as well as the classical entropy) for the monitored dynamics.
We observe that in the periodic boundary conditions, the entanglement entropy scaling indeed undergoes a logarithm-to-area-law transition, while this transition is missing in the open boundary conditions.
Also, to show the effect of the trajectory variance, as well as find a tighter bound on the entanglement entropy, we calculate the $\overline{S_\text{cl}[\{\langle n_i \rangle\}]}$.

In the above analysis, we consider only the $\theta = \pi$ case in the feedback operator, while similar FISE appears for arbitrary $\theta$.
In Appendix~\ref{different-theta}, we show the numerical results for $\theta \ne \pi$.
The length of the domain wall, however, is minimized at $\theta = \pi$.
When $\theta$ approaches zero, we expect that the FISE still appears, although with a large domain wall that may exceed the numerical simulation capability.
Therefore, for the projector $P_i$, the FISE is not a fine-tuned phenomenon and may appear for a large class of feedback operations.

\subsection{Bulk-edge correspondence}

The bulk-edge correspondence of the non-Hermitian effect has been extensively investigated \cite{NHSE-3,NHSE-4,NHSE-5,NHSE-6,NHSE-7}.
In the no-click limit, where the dynamics are described by the non-interacting non-Hermitian Hamiltonian, the skin effect also manifests as a directional bulk current in the periodic boundary systems:
\begin{equation}
	J[\{n_k\}] \equiv \int_{-\pi}^{+\pi} v_k n_k dk,\quad
	v_k = \partial_k E(k)
\end{equation}
where $n_k =\langle c_k^\dagger c_k \rangle$ is the distribution function of the $k$-momentum fermion modes, and $E(k)$ is the dispersion of $H_\text{eff}$. 

The rigorous bulk-edge correspondence only happens in the no-click limit, whereas when the quantum jump processes are taken into account, the system loses the single-particle description, and thus the velocity $v_k$ becomes ill-defined.
Nonetheless, as our primary focus lies in the asymptotic behavior as $\gamma$ approaches zero, we can extend the notion of the directional bulk current $J[{n_k}]$. 
This extension involves an approximation where the velocity $v_k$ is substituted with the outcome of the nonmonitored free system, denoted as 
\begin{equation}
	v_k \approx \partial_k E_o(k) = \sin(k).
\end{equation}
Here, $E_o(k)$ is the dispersion of the nonmonitored Hamiltonian $H$.

By simulating the system with identical parameters but under periodic boundary conditions, we show in Fig.~\ref{Fig}(e) that there is an imbalance in the momentum distribution which does not vanish in the $\gamma \rightarrow 0$ limit.
Specifically, for $\gamma=0.01$ case, the numerical simulation shows that $J[\{n_k\}] \approx -0.94$.
This nonzero current serves as another manifestation of the dynamical skin effect at arbitrarily small measurement rates.

\subsection{Monitored interacting system}

The monitored free-fermion systems belong to a special type of MIET since the entanglement of the weakly monitored system does not follow the volume-law scaling.
From the quantum information point of view, the free fermion systems are not good quantum memory; the initial quantum information will be lost in polynomial time when subject to measurements \cite{Hastings2021}.
It is therefore necessary to exclude the possibility that FISE only suppresses the logarithm-scaling entanglement.

For the interacting system, it is more convenient to focus on a spin-1/2 Hamiltonian
\begin{equation}\label{eq:int-model}
\begin{aligned}
	H =&\ \sum_i \left[J_1(\sigma_i^x \sigma^x_{i+1} + \sigma_i^y \sigma^y_{i+1})+  J_z \sigma_i^z \sigma^z_{i+1}\right.  \\
	&\ + \left.J_2 (\sigma_i^x \sigma^x_{i+2} + \sigma_i^y \sigma^y_{i+2})\right].
\end{aligned}
\end{equation}
In Ref.~\cite{Fuji-2020}, the nonintegrable system (\ref{eq:int-model}) under continuous monitoring (on local $\sigma_i^z$) was shown to display a volume-to-area-law entanglement transition.

In our case, however, we choose the same generalized monitoring as in the free fermion case (via Jordan-Wiger transformation):
\begin{equation}
	P_i = \frac{1}{2}(\sigma_i^+ -i\sigma_{i+1}^+)(\sigma_i^- + i\sigma_{i+1}^-),\quad U_i = S^z_{i+1}.
\end{equation} 
In the presence of interactions, FISE also manifests in open boundary systems under the influence of this generalized monitoring. 
The presence of the dynamical skin effect leads to area-law entanglement entropies, allowing for efficient representation of states using matrix-product states \cite{mps-1,mps-2}. 
Additionally, the time evolution of large system sizes can be simulated effectively using the TEBD algorithm \cite{TEBD-1, TEBD-2}. 

We investigate the steady-state classical entropy for different $\gamma$ values and system sizes up to $L=64$, and fix $J_1 = 0.5$, $J_z=1.0$, $J_2=0.1$. 
As shown in Fig.~\ref{Fig}(f), the numerics indicate that for $\gamma>0.5$, the classical entropy exhibits the asymptotic behavior: 
\begin{equation}\label{eq:scaling-form-4}
	S_\mathrm{cl}(\gamma, t>t_\text{rlx}, L > L_\text{DW}) \approx \frac{\tilde c}{\gamma^{\nu}},
\end{equation}
where $\tilde c=1.8$ and $\nu=0.85$.
This suggests that a matrix product state with bond dimension $\chi \sim O(\exp(\gamma^{-\nu}))$ is sufficient to describe the state.
However, the numerical simulation becomes challenging in the small-$\gamma$ regime. 
While we are unable to numerically verify the scaling behavior in the small-$\gamma$ limit due to computational limitations, our finite-size simulations still demonstrate the scaling law described by [as depicted in Fig.~\ref{Fig}(f)]
\begin{equation}\label{eq:int-scale}
	S_\text{cl}(\gamma, t > t_\text{rlx}) = L \cdot \tilde G (\gamma L^{1/\nu}),
\end{equation}
where the scaling function shall obey the asymptotic behavior
\begin{equation}
	\tilde G (x \rightarrow \infty) \sim \frac{\tilde c}{x^\nu}.
\end{equation}
We conjecture that the asymptotic form in Eq.~(\ref{eq:int-scale}) continues to hold in the small-$\gamma$ regime.

\begin{figure*}
	\centering
	\begin{tabular}{cc} 
        (a) Entanglement Entropy scaling & (b) Mutual Information \\
        \includegraphics[height=48mm]{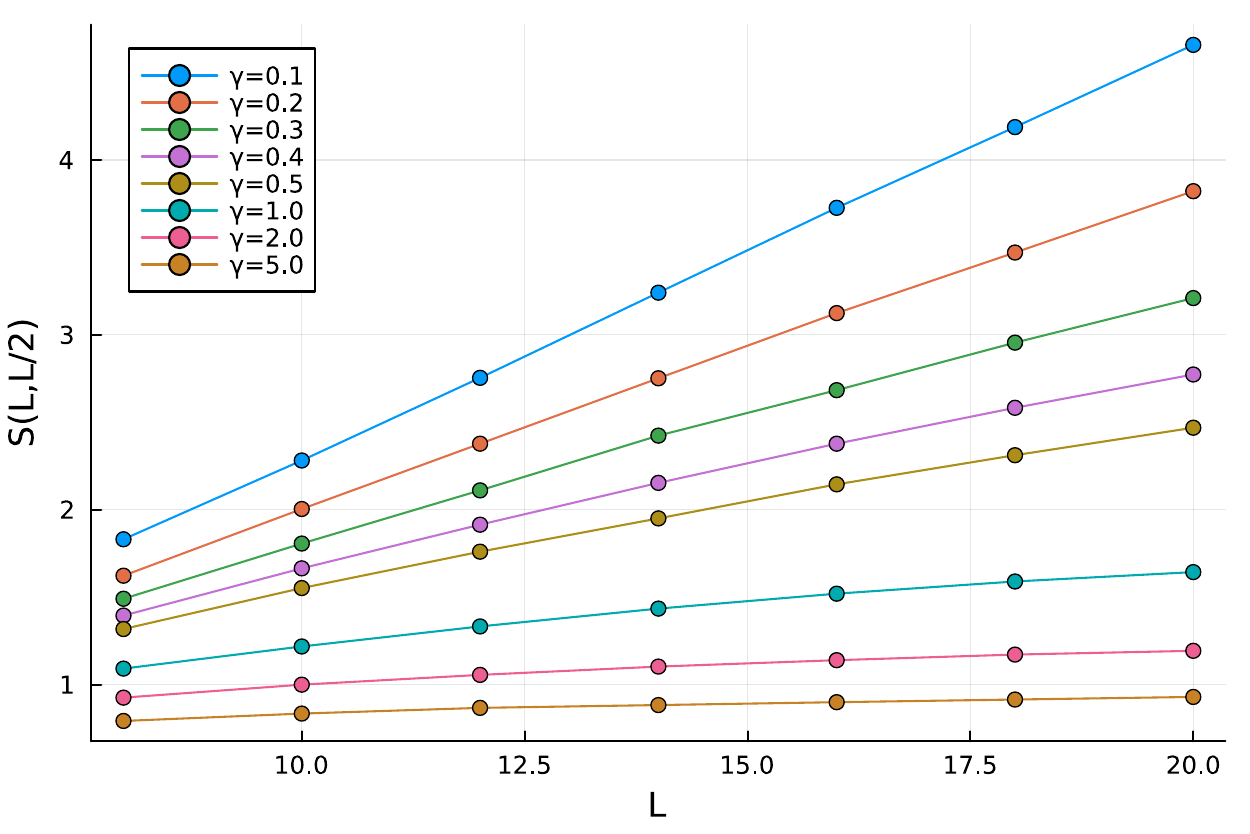} & 
        \includegraphics[height=48mm]{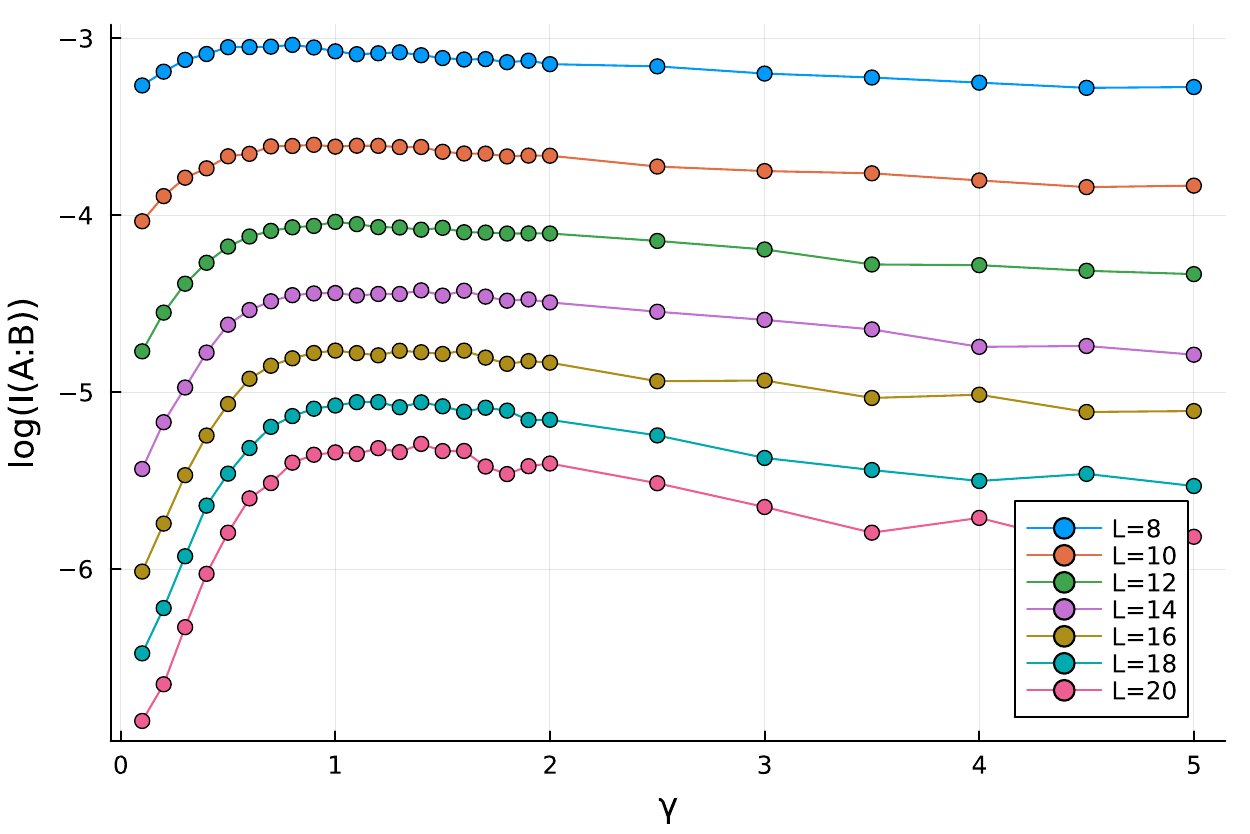}
        \end{tabular}
	\caption{Exact diagonalization simulation of the entanglement dynamics for systems with $J_1=J_z=0.5$, $J_2=0.1$, and under periodic boundary conditions. (a) Trajectory average of entanglement entropy for different measurement rates. (b) Trajectory average of mutual information $I(A:B)$ between two antipodal sites $A$ and $B$. The peak of $I(A:B)$ indicates the critical measurement rate at $\gamma_c \approx 1$.}
\label{fig:C5-2}
\end{figure*}

To complete the analysis, we show that under periodic boundary conditions, the usually measurement-induced entanglement appears in this generalized monitored system.
We demonstrate this transition using finite-size exact diagonalization.
We remark here that the appearance of an entanglement transition can be better revealed from the mutual information $I(A:B)$ shared by two antipodal sites $A$ and $B$.
This numerical technique has been used, for example, in Refs.~\cite{PhysRevB.100.134306,Fuji-2020}.

In Fig.~\ref{fig:C5-2}(a), we first demonstrate that for small $\gamma$, the finite-size scaling of entanglement entropy $S(L,L/2)$ shows a linear growth in the size $L$.
In the large $\gamma$ regime, the entanglement is greatly suppressed, suggesting steady states being area-law entangled.
To better demonstrate the existence of an entanglement transition, Fig.~\ref{fig:C5-2}(b) displays the mutual information $I(A:B)$.
The peak in $I(A:B)$ shows near $\gamma_c \approx 1$.
Due to the finite size available to the numerical simulation, we are unable to pinpoint the critical measurement rate $\gamma_c$.

Therefore, we show that the FISE indeed imposes strong suppression on the entanglement which completely excludes any entanglement transition.

\section{Floquet circuits under random measurements}
\label{floquet}
 
Till now, we have investigated FISE in the context of the continuously monitored Hamiltonian dynamics.
Readers may wonder whether FISE is an artifact of continuous monitoring.
This section serves to clarify this question by generalizing the FISE to the quantum circuits under random (generalized) measurements, which is similar to the original setup for the MIET.
The upshot is that only the discrete spatial-temporal translation symmetry is required for the skin effect.

\begin{figure*}
	\centering
	\includegraphics[width=\linewidth]{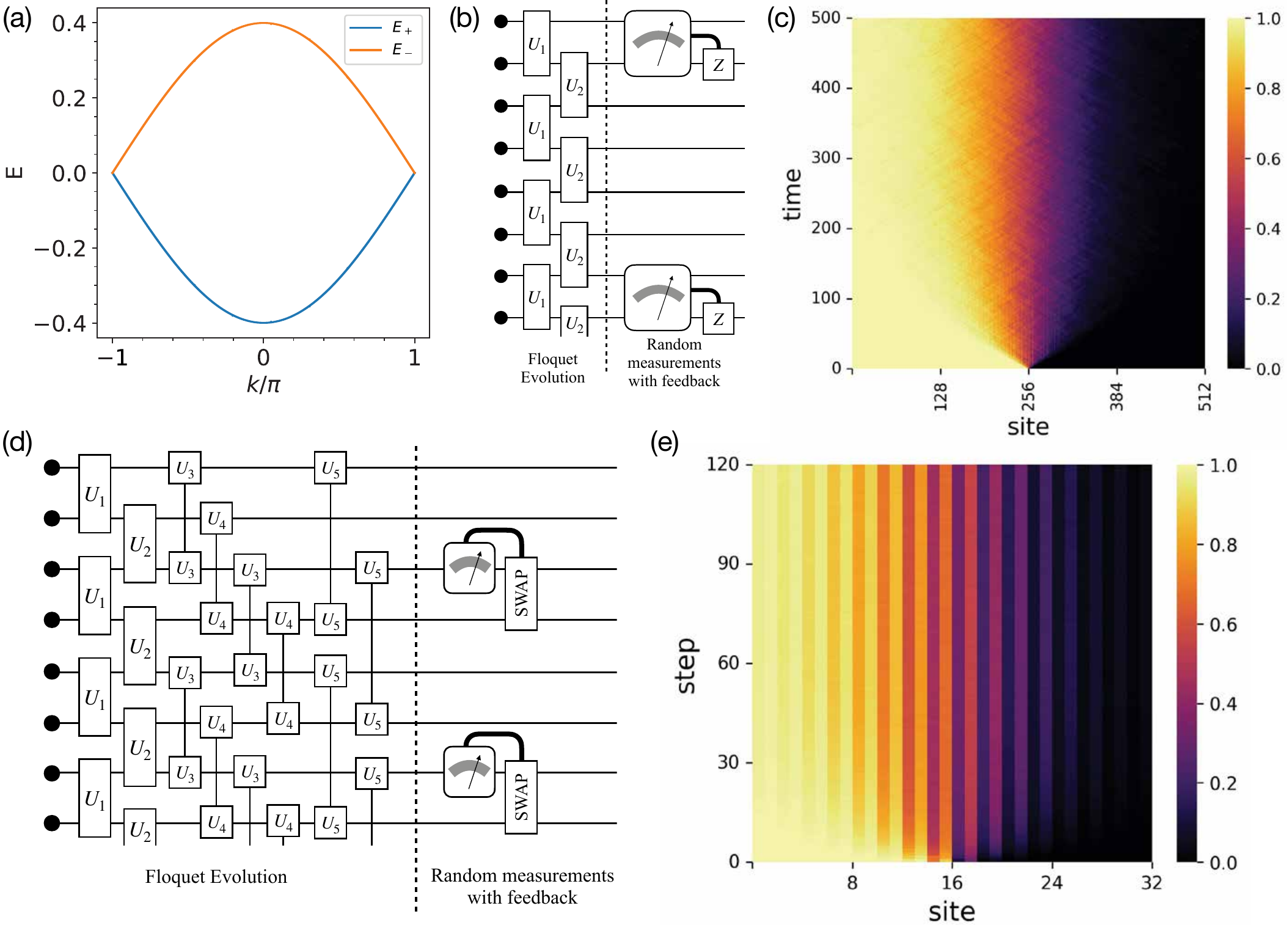}
	\caption{(a) Dispersion relation for circuit model. At the maximal velocity point $k = \pi$, the eigenvectors are $|\pi,-\rangle \approx 0.842 |\pi,1\rangle + (0.290-0.455i)|\pi,2\rangle$ and $|\pi,+\rangle \approx (0.290+0.455i) |\pi,1\rangle - 0.842|\pi,2\rangle$. (b) Quantum circuit model with Floquet circuit evolution and conditional feedback. $U_1$ and $U_2$ are defined as Eq.~\ref{eq:floquetU}. Random measurements are defined as $P_i$ Eq.~\ref{eq:floquetM}. Feedback $\sigma_i^z$ follows the measurement $P_i$ if the measurement outcome is $1$. (c) Density evolution of the measured Floquet dynamics ($L=512$, $p=0.02$). (d) Experimentally realizable quantum circuit model designed for trapped ions quantum computers. Unitary gates $U_1\sim U_5$ are defined as Eq.~\ref{eq:experimentU}. Measurements are simple onsite measurements $\sigma_{2i-1}^z$ on odd sites. Feedback $\operatorname{SWAP}_{2i-1,2i}$ follows the measurement $\sigma_{2i-1}^z$ if the measurement outcome is $-1$. (e) Classical simulation of the density evolution on a 32-qubit trapped ion chain ($\theta = \pi/4$, $p=0.5$).}
	\label{fig:floquet}
\end{figure*}

\subsection{Wave packet motion under Floquet evolution}

The notion of discrete quantum circuits naturally appears when we use the Trotter decomposition.
In general, a local Hamiltonian can be decomposed into several groups:
\begin{equation}
	H = \sum_{\alpha=1}^l H_\alpha, \quad H_\alpha = \sum_i o_{\alpha,i},
\end{equation}
where each $o_{\alpha,i}$ is a local operator.
Within each group, the local operators do not overlap.
The time evolution can then be approximated by
\begin{equation}\label{eq:trotter}
	U(t=N\Delta t) \approx \left( e^{-i H_1 \Delta t} e^{-i H_2 \Delta t}\cdots e^{-iH_l \Delta t} \right)^N.
\end{equation}
The decomposition approaches the real dynamics in the $\Delta t \rightarrow 0$ limit.
While for practical reasons, $\Delta t$ is a finite value, therefore the dynamics described by Eq.~(\ref{eq:trotter}) is a Floquet dynamics with period $T = l\Delta t$.

We show that the notion of wave packet motion is also valid in the context of the Floquet circuit even when the $\Delta t$ is not vanishing small, as long as the discrete translational symmetry is preserved.
For simplicity, we assume that the Floquet circuit has two-site translational symmetry.
We can group two lattice sites into a unit cell and label them as the internal degrees of freedom $a=1,2$.
The plane wave is then
\begin{equation}
	|k,a\rangle \equiv \frac{1}{\sqrt{L}}\sum_j c_{j,a}^\dagger e^{-ikj}|0\rangle.
\end{equation}
Due to the translational invariance, the evolution $U(T)$ is block diagonal in the momentum space:
\begin{equation}
\begin{aligned}
	U_{ab} &\equiv \langle k,a|U(T)|k,b\rangle \\
	&=V(k) \begin{bmatrix}
		e^{-iE_+(k) T} & 0 \\ 0 & e^{-iE_-(k) T}
	\end{bmatrix} V(k)^\dagger.
\end{aligned}
\end{equation}
The values $E_\pm(k)$ extracted from the eigenvalues of $U_{ab}$ are the quasi-energies of the Floquet dynamics.
Consider the branch with quasi-energy $E_+$.
The wave packet with averaged momentum $k$ and averaged position $x$ has the form
\begin{equation}
	|k,x;+\rangle \equiv \int \frac{dp}{2\pi} e^{-\alpha (p-k)^2 - ix(p-k)} |p,+\rangle,
\end{equation}
where $\alpha$ controls the variance of the momentum distribution of the wave packet.
After one period $T$, the wave packet evolves to
\begin{equation}
\begin{aligned}
	&\ U(T)|k,x;+\rangle \\
	=& \int \frac{dp}{2\pi} e^{-\alpha (p-k)^2 - ix(p-k) - iE_+(p)T} |p,+\rangle \\
	\propto& \int \frac{dp}{2\pi} e^{-\alpha (p-k)^2 - i\left[x+\frac{E_+(p)-E_+(k)}{p-k}T\right](p-k)} |p,+\rangle.
\end{aligned}
\end{equation}
Specifically, when $\alpha$ is large, most $p$ will be close to $k$, and the movement of the averaged position is approximated by $x \rightarrow x^\prime \approx x + E^\prime_{+}(k) T$.
We therefore restore the wave packet moving picture in the Hamiltonian dynamics, at least for the nearly monochromatic wave packets.

With the picture in mind, we first consider the Floquet dynamics described by the circuits
\begin{equation}\label{eq:floquet_1}
\begin{aligned}
	U(T) =&\ \prod_j\exp\left[-i \left(c_{2j}^\dagger c_{2j+1}+c_{2j+1}^\dagger c_{2j}\right)T \right] \\
	  &\times \prod_j\exp\left[-i \left(c_{2j-1}^\dagger c_{2j}+c_{2j}^\dagger c_{2j-1}\right)T \right].
\end{aligned}
\end{equation}
We can choose an arbitrary finite value of $T$, say $T=0.5$. 
The dispersion is then calculated and displayed in Fig.~\ref{fig:floquet}(a).
We remark that for finite $T$, the eigenvectors have shifted compared to the Hamiltonian case.
However, the tendency remains the same.
That is, the configuration $d^\dagger |0\rangle$ still has a bigger overlap with the right-moving wave packet than with the left-moving ones.
Therefore, we expect that the feedback-induced skin effect still appears in the Floquet model.

\subsection{Skin effect in measured Floquet circuits}

Besides the generalization to the Floquet circuit, we show here that the continuous monitoring setup can be replaced with randomly applying strong projective measurement followed by conditional feedback.

Consider the circuit dynamics in Fig.~\ref{fig:floquet}(b), where the $U_1$ and $U_2$ come from the Floquet dynamics.
Under the Jordan-Wigner transformation, they become
\begin{equation}\label{eq:floquetU}
	U_1 = U_2 = \exp\left(-i\frac{T}{2}\sigma_x\otimes\sigma_x\right)\exp\left(-i\frac{T}{2}\sigma_y\otimes\sigma_y\right),
\end{equation}
and the measurements are the projective measurements of the observable
\begin{equation}\label{eq:floquetM}
	P_i = \frac{1}{4}\left(\sigma^z_i + \sigma^z_{i+1} + \sigma^x_i\sigma^y_{i+1}-\sigma_i^y\sigma^x_{i+1}+2\right),
\end{equation}
followed by the conditional feedback $U_i = \sigma_i^z$.
As displayed in Fig.~\ref{fig:floquet}(c), for a small measurement rate ($p=0.02$), the feedback still induces the skin effect.

\subsection{Experimental proposal on the trapped ions quantum computers}

The two-site measurements are hard to realize in the experiments.
However, we can devise a single-site measured model using a unitary transformation.
To formulate the transformation more clearly, we temporarily reverse to the fermionic model where the quasi-mode $d_i = c_i + i c_{i+1}$ are being measured.
We now regard $d_{2i-1}$ as the local fermion mode, which corresponds to the transformation
\begin{equation}
\begin{aligned}
	d_{2i-1} &= \frac{1}{\sqrt 2}(c_{2i-1} + i c_{2i}), \\
	d_{2i} &= \frac{1}{\sqrt 2}(c_{2i} + i c_{2i-1}).
\end{aligned}
\end{equation}
The Hamiltonian is mapped to $H = h_1 + h_2 + h_3 + h_4$, where $\{h_1,h_2,h_3,h_4\}$ are hopping terms among neighboring unit cells:
\begin{equation}\label{eq:hopping-ham}
\begin{aligned}
	h_1 &= \sum_i \left(d_{2i-1}^\dagger d_{2i} + H.c. \right), \\
	h_2 &= \frac{1}{2} \sum_i \left(d_{2i}^\dagger d_{2i+1} + H.c. \right), \\
	h_3 &=  \sum_i \left(\frac{i}{2}d_{2i-1}^\dagger d_{2i+1} - \frac{i}{2}d_{2i}^\dagger d_{2i+2} + H.c. \right), \\
	h_4 &= \frac{1}{2} \sum_i \left(d_{2i-1}^\dagger d_{2i+2} + H.c. \right).
\end{aligned}
\end{equation}
Under the new basis, the projector becomes an onsite measurement of particle occupation on the odd sites:
\begin{equation}
	P_{2i-1} = d_{2i-1}^\dagger d_{2i-1},\quad P_{2i} = d_{2i} d^\dagger_{2i}.
\end{equation}
The feedback operator, however, is not in a neat form on this basis.
However, as we remarked, the FISE is not a fine-tuned effect from the feedback, we expect FISE to happen for a wide range of conditional unitary operators.
For simplicity, we choose the feedback to be the swap operator on two neighboring sites.
To obtain an even simpler model, we keep only half of the measurement $P_{2i-1}$ on the odd sites, and we choose the feedback operator to be the swap gate 
\begin{equation}
	U_{2i-1} = \operatorname{SWAP}_{2i-1,2i}.
\end{equation}

The fermion model can be mapped to a spin model using the Jordan-Wigner transformation. 
The hopping terms beyond the nearest neighbor will take the form of string operators, which is hard to implement. 
However, the numerics show that even if we omit the Jordan-Wigner string, the resulting model still shows FISE.
In this setup, the measured circuit is displayed in Fig.~\ref{fig:floquet}(d).
The unitary gate $\{U_1,\cdots,U_5\}$ in the circuits are: 
\begin{equation}
\begin{aligned}\label{eq:experimentU}
	U_1 &= \mathrm{XX}(\theta) \otimes \mathrm{YY}(\theta), \\
	U_2 &= \mathrm{XX}\left(\frac{\theta}{2}\right) \otimes \mathrm{YY}\left(\frac{\theta}{2}\right), \\
	U_3 &= \mathrm{XY}\left(\frac{\theta}{2}\right) \otimes \mathrm{YX}\left(-\frac{\theta}{2}\right), \\
	U_4 &= \mathrm{XY}\left(-\frac{\theta}{2}\right) \otimes \mathrm{YX}\left(\frac{\theta}{2}\right), \\
	U_5 &= \mathrm{XX}\left(\frac{\theta}{2}\right) \otimes \mathrm{YY}\left(\frac{\theta}{2}\right),
\end{aligned}
\end{equation}
where $\theta$ is the parameter controlled by the time step of the Trotter decomposition, and the $\mathrm{XX}$, $\mathrm{XY}$, and $\mathrm{YY}$ gates are
\begin{eqnarray}
	\mathrm{XX}_{i,j}(\theta) &\equiv &\exp\left(-i\theta \sigma^x_i \sigma^x_j \right), \\
	\mathrm{XY}_{i,j}(\theta) &\equiv &\exp\left(-i\theta \sigma^x_i \sigma^y_j \right), \\
	\mathrm{YY}_{i,j}(\theta) &\equiv & \exp\left(-i\theta \sigma^y_i \sigma^y_j \right).
\end{eqnarray}
The measurement operator is now simply the $\sigma^z_{2i-1}$ on odd sites, and the conditional feedback $\operatorname{SWAP}$ (which does not change under Jordan-Wigner transformation) will be applied to the sites ($2i-1$) and ($2i$) when the measurement result in site ($2i-1$) is $-1$.

Experimentally, the trapped ions systems \cite{RMP.62.531,PRL.74.4091,PRL.104.140501,PhysRevLett.82.1835} are one of the ideal platforms to realize medium-scale quantum computation and quantum simulations.
Using the ${}^{171}\mathrm{Yb}^+$ ions trapped by the high-frequency electromagnetic field \cite{RMP.62.531} with fined tune laser pulse, the platform is able to:
\begin{enumerate}
	\item initiate the state as $|0\cdots0\rangle$.
	\item implement any single-site rotation, which enables initiation of the state to any product state.
	\item implement an ``Ising" spin-spin interaction \cite{PRL.92.207901,S.L.Zhu.EPL.2006}: $\mathrm{XX}_{i,j}(\theta)=\exp\left(-i\theta \sigma^x_i \otimes \sigma^x_j\right)$.
	\item do onsite measurement on $\sigma^z$, collapsing qubit to $|0\rangle$ or $|1\rangle$ state.
\end{enumerate}
We remark that using the combination of single-site and two-site unitary gates, it is possible to realize the YY and XY gates as:
\begin{equation}
\begin{aligned}
	\mathrm{XY}_{i,j}(\theta) &= \exp\left(-i\frac{\pi}{2}\sigma^z_j\right) \mathrm{XX}_{i,j}(\theta) \exp\left(i\frac{\pi}{2}\sigma^z_j\right), \\
	\mathrm{YY}_{i,j}(\theta) &= \exp\left(-i\frac{\pi}{2}\sigma^z_i\right) \mathrm{XY}_{i,j}(\theta) \exp\left(i\frac{\pi}{2}\sigma^z_i\right).
\end{aligned}
\end{equation}

Now consider the monitored circuit model displayed in Fig.~\ref{fig:floquet}(d), where the Floquet dynamics is the Trotterized version of Hamiltonian (\ref{eq:hopping-ham}).

The measurement that needs to perform the simple $S^z$ measurement, followed by a conditional swap operation, which is
\begin{equation}\label{eq:experimentF}
	U_\text{SWAP} \propto \mathrm{XX}\left(\frac{\pi}{4}\right)\otimes \mathrm{YY}\left(\frac{\pi}{4}\right). 
\end{equation}
In Fig.~\ref{fig:floquet}(e), we simulate the dynamics on a $32$-qubit chain, and choose $\theta=\pi/4$, with measurement probability $p=0.5$.
We see the late time state show a clear feedback-induced skin effect.

\section{Conclusion}
\label{conclusion}
This work investigates the effect of generalized monitoring on the entanglement phase transition, with a particular focus on the emergence of the skin effect in certain open boundary monitored systems. 
Our analysis reveals that the skin effect qualitatively alters the entanglement structure of the nonequilibrium steady state, leading to a single area-law phase. 
Specifically, we show that introducing generic feedback operators can disturb the balance of particle distribution, resulting in particle accumulation. 
We demonstrate that the FISE is not a fine-tuned phenomenon, as the skin effect appears for different feedback parameters, and survives in the presence of interactions.
The suppression of the entanglement entropy from the skin effect also enables an efficient classical simulation of the monitored interacting systems.

These results have practical implications in the context of open systems or controllable quantum devices, as the monitoring-feedback setup can enable the realization of a skin effect without the need for post-selection in non-Hermitian dynamics. 
For systems showing FISE, the steady states can be reached in constant steps, which is accessible for the noisy intermediate-scale quantum devices.
We thus proposed a quantum circuit model displaying FISE which can be experimentally realized on the trapped ions systems.

\section{Note added}
In the middle of this work, we became aware of a recent work \cite{2208.10506}, which also considers the effect of generalized monitoring in the context of MIET.
The two works are complementary to each other: in Ref.~\cite{2208.10506}, the authors utilize the feedback (pre-selection) to reveal MIET as a quantum absorbing state transition that can be directly detected, while our work shows that the presence of conditional feedback may also eliminate the MIET.

\begin{acknowledgements}
J.R. thanks Chenguang Liang for the valuable discussions. 
J.R. and Y.P.W. thank Xu Feng and Shuo Liu for alerting them of a typo in the previous version.
The numerical simulations based on tensor-network use the \texttt{ITensor} package \cite{itensor}. 
\end{acknowledgements}

\begin{appendix}

\section{Uniqueness of nonequilibrium steady state}
\label{uniqueness}
In this section, we prove the uniqueness of $\rho_\mathrm{NESS}$ for models considered in the main text, by explicitly checking the operator set
\begin{equation}
	\mathcal O = \{H, L_1, L_1^\dagger,L_2,L_2^\dagger,\cdots\}
\end{equation}
generates (under multiplication and addition) the complete algebra on the Hilbert space.

\subsection{Projective monitoring}
We first prove that the Hamiltonian (under open boundary conditions)
\begin{equation}
	H = \sum_i(c_i^\dagger c_{i+1} + c_{i+1}^\dagger c_i)
\end{equation}
and the projectors
\begin{equation}
	P_i = \frac{1}{2}(c_i^\dagger -i c_{i+1}^\dagger) (c_i+i c_{i+1})
\end{equation}
generate the whole algebra.
Note $H$ and $P_1$, $P_2$ together generate the following particle number operators:
\begin{equation}\label{apx:comm-1}
\begin{aligned}
	n_1 - n_3 &= P_2 - P_1 + i[H,P_1+P_2], \\
	n_1 + n_2 &= \frac{1}{2}\left(P_1+P_2+i[H,n_1-n_3]+n_1-n_3\right), \\
	n_2 + n_3 &= (n_1 + n_2) - (n_1 - n_3).
\end{aligned}
\end{equation}
Then, some straightforward algebra leads to
\begin{equation}\label{apx:comm-2}
\begin{aligned}
	c_1^\dagger c_2 - c_2^\dagger c_1 &= i (n_1+n_2) - i P_1, \\
	c_1^\dagger c_2 + c_2^\dagger c_1 &= [c_1^\dagger c_2 - c_2^\dagger c_1, n_2+n_3], \\
	c_2^\dagger c_3 - c_3^\dagger c_2 &= i (n_2+n_3) - i P_2, \\
	c_2^\dagger c_3 + c_3^\dagger c_2 &= [n_1+n_2,c_2^\dagger c_3 - c_3^\dagger c_2].
\end{aligned}
\end{equation}
Upon some addition among Eqs.~(\ref{apx:comm-2}), we obtain the operator $c_1^\dagger c_2$, $c_2^\dagger c_3$ and their Hermitian conjugates.
The commutations of them further produce $c_1^\dagger c_3$ and its conjugate.
Also, note that
\begin{equation}
	[c_1^\dagger c_2, c_2^\dagger c_1] = n_1-n_2.
\end{equation}
Together with Eqs.~(\ref{apx:comm-1}), we generate all fermion bilinear terms $c_i^\dagger c_j$ (including $i=j$ case) on sites $(1,2,3)$.
To proceed, we subtract the hopping terms between sites $(1, 2)$ from $H$.
The resulting operator is equivalent to a shorter chain starting from site 2.
We can then utilize the calculation above to obtain all $c_i^\dagger c_j$ terms on sites $(2,3,4)$.
We eventually obtain all fermion bilinear terms on the chain by applying the strategy iteratively.
Note that fermion bilinear terms $c_i^\dagger c_j$ generate the complete algebra within a fixed particle-number sector since any two product states in the sector can be related by applying several fermion hopping terms.

\subsection{Generalized monitoring and interactions}
For the generalized monitored system described by the jump operators $\{L_n = U_n P_n\}$, the proof of uniqueness is essentially the same as the projective case.
Note that we can generate all $P_i$ terms by multiplying to jump operators 
\begin{equation}
	P_i = L_i^\dagger L_i.
\end{equation}
In this way, we can generate the complete operator algebra in the same way as above.

We argue that the completeness of the operator algebra holds for generic open systems since the exact decoupling of Hilbert space is the result of symmetries or fine-tuning.
For the interacting system where the Hamiltonian is
\begin{equation}
	H = \sum_i(c_i^\dagger c_{i+1} + c_{i+1}^\dagger c_i + gn_i n_{i+1}).
\end{equation}
The above argument means for a random value of $g$, we should expect the completeness of the operator algebra.
We can also consider the case where the coupling constant is smoothly varying in the space.
In particular, let $g_i = 0$ for the sites near the boundary.
Following the same procedure, we can generate the operator algebra of the subsystem near the boundary.
Assume that we meet the first nonzero $g$ at site $i+1$.
It means that we have the complete operator algebra (within fixed filling number) $\mathcal A_{[1,i]}$ of the subsystem consisting of sites $1,\dots,i$.
We first subtract all terms within $\mathcal A_{[1,i]}$ from the Hamiltonian and denote the result as $H_{[i+1,N]}$.
Consider the commutators
\begin{equation}
\begin{aligned}
	[n_{i-1}, H_{[i+1,N]}] &= c_{i}^\dagger c_{i+1} - c_{i+1}^\dagger c_{i}, \\
	[n_{i-1}, c_{i}^\dagger c_{i+1} - c_{i+1}^\dagger c_{i}] &= c_{i}^\dagger c_{i+1} + c_{i+1}^\dagger c_{i}.
\end{aligned}
\end{equation}
In this way, we generate the hopping terms $c_{i}^\dagger c_{i+1}$ and $c_{i+1}^\dagger c_{i}$.
Those terms together with $\mathcal A_{[1,i]}$ generate the algebra $\mathcal A_{[1,i+1]}$.
This procedure can proceed iteratively, therefore producing the complete algebra.

\section{Stochastic Schr\"{o}dinger equation}

\subsection{Generalized monitoring}
\label{generalized measurement}

Microscopically, a measurement process involves a short-time interaction between the system and the probe, which are initially separable:
\begin{equation}\label{apx:sys-probe}
	|\psi_{AB}\rangle = e^{-i H_\mathrm{int} \Delta t} |\psi_A\rangle \otimes |\psi_B\rangle,
\end{equation}
where the wave function of the measured system is denoted as $|\psi_A\rangle$ and the probe $|\psi_B\rangle$.
When $\Delta t$ is much smaller than the time scale of the system, the system can be regarded as static during the measurement.
Such measurement is called the \textit{strong measurement}.
The probe is thought to be a device that can convert quantum information to the classical one, which takes the form of standard projective measurement.
That is, suppose the eigenbasis of the probe is $\{|\phi_n\rangle\}$, the probability of getting a record $n$ is
\begin{equation}
	p_n = \langle \phi_n| \rho_B |\phi_n\rangle,\quad 
	\rho_B \equiv \operatorname{Tr}_A |\psi_{AB}\rangle\langle \psi_{AB}|,
\end{equation}
and the feedback of the measurement to the system is
\begin{equation}
	|\tilde\psi_A^{(n)}\rangle = \langle\phi_n| e^{-i H_\mathrm{int} \Delta t} |\psi_A\rangle \otimes |\psi_B\rangle \equiv M_n |\psi_A\rangle.
\end{equation}
The completeness condition requires 
\begin{equation}
	\sum_n \langle\tilde\psi_A^{(n)}|\tilde\psi_A^{(n)}\rangle=1 
	\ \Longrightarrow \ 
	\sum_n M_n^\dagger M_n = 1.
\end{equation}
This is the general form of the measure.
In the language of density operator, a measurement is described by a set of operators $\{M_n\}$.
A measurement process may record a result $n$ with probability $p_n$ and change the state to:
\begin{equation}
	\rho \rightarrow \frac{M_n \rho M_n^\dagger}{\|M_n \rho M_n^\dagger\|}.
\end{equation}
If the measurement result is not known, the averaged density matrix after the measurement is
\begin{equation}
	\rho \rightarrow \sum_n M_n \rho M_n^\dagger.
\end{equation}
Such a map is called the quantum channel \cite{nielsen_chuang_2010}.

On the other hand, if the strength of system-probe coupling is comparable with the energy scale of the system, which is the case for an open quantum system, the quantum channel expression should depend on time $\Delta t$.
This kind of measurement process is called \textit{weak measurement}.
When the system is \textit{Markovian} (the equation of motion depends only on the near past), the course-grained dissipation process can be described by the channel:
\begin{eqnarray}
	M_n &=& L_n \sqrt{\gamma \Delta t}, \\
	M_0 &=& \sqrt{1 - \sum_{n>0} M_n^\dagger M_n} \nonumber \\
	&=& 1-\frac{\gamma}{2}\sum_{n>0} L_n^\dagger L_n \Delta t + O(\Delta t^2).
\end{eqnarray}
For the density matrix, the coarse-grained differential equation is the Lindblad equation:
\begin{equation}
\begin{aligned}
	\frac{d\rho}{dt} &= \lim_{\Delta t \rightarrow 0} \frac{1}{\Delta t}\sum_n M_n e^{-iH\Delta t}\rho e^{iH\Delta t} M_n^\dagger \\
	&= -i [H,\rho]-\frac{\gamma}{2}\sum_{i} \{L_i^\dagger L_i, \rho\} + \gamma\sum_{i} L_i\rho L_i^\dagger.
\end{aligned}
\end{equation}

\begin{figure}
	\centering
	\includegraphics[width=0.95\linewidth]{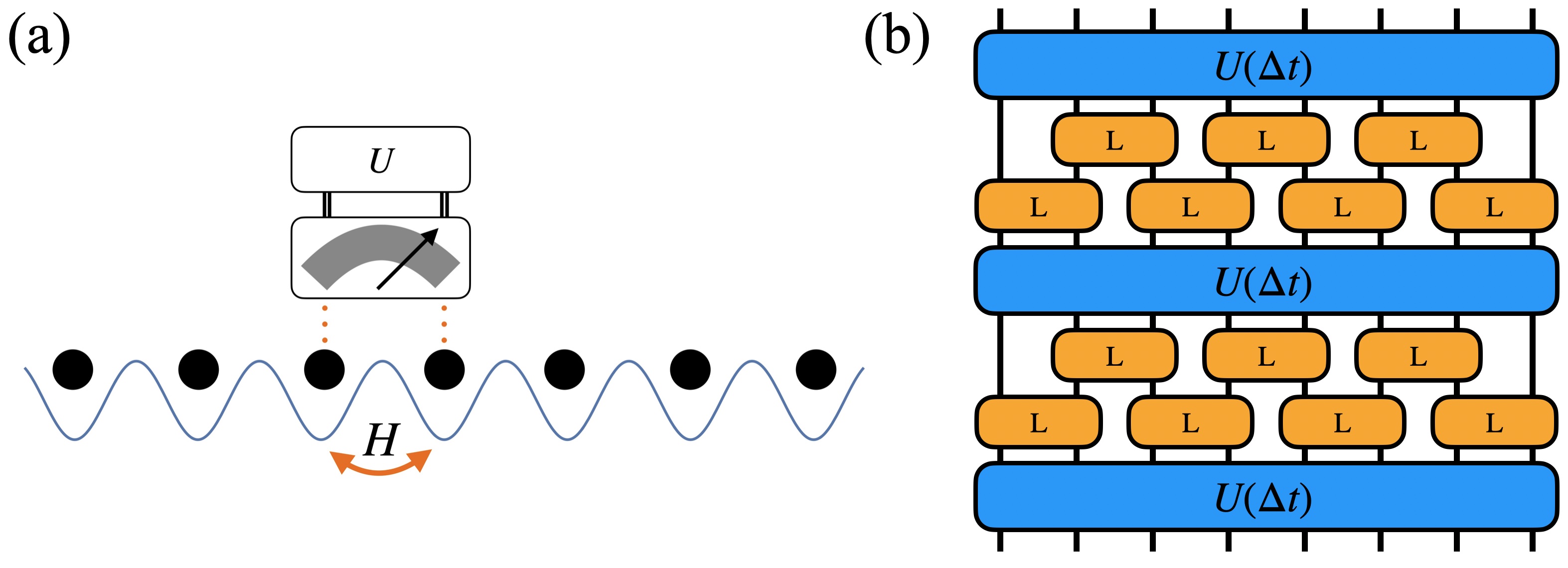}
	\caption{(a) Schematic of a fermion chain under nearest neighbor interaction and generalized monitoring on each pair of neighboring sites. Unitary feedback will be applied if the probe records a quantum jump. (b) Circuits representation of the discretized Hamiltonian evolution with constant monitoring.}
	\label{apx:circuits-fig}
\end{figure}

The joint dynamics of Hamiltonian evolution and measurement can be equivalently described by the stochastic process, as shown in Fig.~\ref{apx:circuits-fig}, where for each time step $\Delta t$, the system first undergoes a coherent evolution $|\psi\rangle \rightarrow e^{-iH\Delta t}|\psi\rangle$, then the application of measurement produces a random process:
\begin{equation}
	|\psi\rangle \rightarrow 
	\begin{cases}
		M_n(\Delta t) |\psi\rangle & P_n=\langle\psi|L_n^\dagger L_n|\psi\rangle \gamma \Delta t \\
		M_0(\Delta t) |\psi\rangle & P_0= 1- \sum_{n>0} P_n
	\end{cases}.
\end{equation}
Different records of the measurement result correspond to different trajectories, and the Lindblad equation is equivalent to the trajectory averaged of such stochastic processes.
In the continuum limit, the stochastic differential equation can be formulated by introducing a Poisson random variable $dW_n$ taking the discrete values of $0$ or $1$.
The $dW_n = 1$ case corresponds to registering a quantum jump, otherwise, $dW_n = 0$.
The expectation value for random $dW_n$ is proportional to $dt$:
\begin{equation}
	\overline{dW}_n = \langle \psi| L_n^\dagger L_n|\psi\rangle \gamma dt.
\end{equation}
Different $dW_n$'s are independent, i.e., they satisfy the orthogonal condition 
\begin{equation}
	dW_m dW_n = \delta_{mn} dW_m.
\end{equation}
Therefore, the random quantum jump process $|\psi\rangle \rightarrow M_n|\psi\rangle$ is described by the expression
\begin{equation}
\begin{aligned}
	d|\psi\rangle = \left(\frac{L_n}{\sqrt{\langle L_n^\dagger L_n\rangle}}-1\right)|\psi\rangle dW_n,
\end{aligned}
\end{equation}
and the null-detection case correspond to $|\psi\rangle \rightarrow M_0 |\psi\rangle$, which is described by a non-Hermitian differential equation:
\begin{equation}
	d|\psi\rangle = -\frac{\gamma}{2}\sum_m (L_m^\dagger L_m-\langle L_m^\dagger L_m\rangle) |\psi\rangle dt,
\end{equation}
where the $\langle L_m^\dagger L_m\rangle$ is introduced for the normalization purpose.
Together with the coherent evolution, we obtain the stochastic Schr\"{o}dinger equation in the main text:
\begin{equation}\label{apx:sse}
\begin{aligned}
	d|\psi\rangle =&\ \left[-i H-\frac{\gamma}{2}\sum_m (L_m^\dagger L_m -\langle L_m^\dagger L_m\rangle)\right] |\psi\rangle dt \\
	 &\ + \sum_m \left(\frac{L_m}{\sqrt{\langle L_m^\dagger L_m\rangle}}-1\right)|\psi\rangle dW_m.
\end{aligned}
\end{equation}

\subsection{Free fermion simulation}
\label{free-fermion-simulation}

For numerical simulation of Eq.~(\ref{apx:sse}), we can first discretize the time into small interval $\Delta t$. 
The discrete evolution is then
\begin{equation}\label{apx:sse-trot}
	|\psi(t+\Delta t)\rangle = \mathcal{M}_{\Delta t} [e^{-iH_\mathrm{eff}\Delta t}|\psi(t)\rangle],
\end{equation}
where $\mathcal M_{\Delta t}$ represents the quantum jump that randomly happened in time interval $\Delta t$:
\begin{equation}\label{eq:discrete-jump}
	\mathcal{M}_{\Delta t}[|\psi\rangle] \propto \prod_{m\in I}L_m |\psi\rangle.
\end{equation}
In Eq.~(\ref{eq:discrete-jump}), the set $I$ denotes the random jump processes, which can be obtained by
\begin{equation}
	I = \{n| r_n < \gamma \langle L_n^\dagger L_n\rangle \Delta t\},
\end{equation}
where $\{r_n \in (0,1)\}$ is a set of independent random variables with evenly distributed probability.

The free fermion system can be efficiently represented by the Gaussian state \cite{gaussianstate}.
For a particle number conserving system, the Gaussian state is a quasimode-occupied state, represented by a matrix $B$:
\begin{equation}
	|B\rangle \equiv \prod_{j=1}^N \sum_i B_{ij} c_{i}^\dagger |0\rangle = \bigotimes_{j=1}^N |B_j\rangle,
\end{equation}
where each column $B_j$ is an occupied quasimode.
Note that there is an SU($N$) gauge freedom for the matrix $B$, i.e., 
\begin{equation}
	|B'\rangle = |BU\rangle = |B\rangle,
\end{equation}
where $U$ is an arbitrary SU($N$) matrix.
Such gauge freedom implies that a Gaussian state is entirely specified by the linear subspace spanned by the quasimodes $B_i$'s.

The random Schr\"{o}dinger equation can be Trotterized as Eq.~(\ref{apx:sse-trot}).
Using the Baker-Campbell-Hausdor formula $e^A B e^{-A} = e^{\operatorname{ad}_A} B$, the nonunitary evolution is  

\begin{equation*}
\begin{aligned}
	e^{-i H_\mathrm{eff} \Delta t} |B_t\rangle 
	&= \prod_{j=1}^N \sum_i B_{ij} e^{-i H_\mathrm{eff} \Delta t} c_i^\dagger e^{i H_\mathrm{eff} \Delta t}|0\rangle \\
	&= \prod_{j=1}^N \sum_i B_{ij} e^{-i \Delta t [H_\mathrm{eff}, \cdot]} c_i^\dagger |0\rangle \\
	&= \prod_{j=1}^N \sum_i \sum_k B_{ij} c_k^\dagger \left[e^{-iH_\mathrm{eff}\Delta t}\right]_{ki} |0\rangle \\
	&= \prod_{j=1}^N \sum_k \left[e^{-iH_\mathrm{eff}\Delta t}B_{ij}\right]_{kj} c_k^\dagger |0\rangle \\
	&= |e^{-i H_\mathrm{eff} \Delta t} B_t\rangle.
\end{aligned}
\end{equation*}

That is, the matrix is multiplied by the exponential of the effective non-Hermitian (single-body) Hamiltonian matrix.
Note that the resulting matrix is not orthogonal anymore, while the state is still well-defined by the linear space spanned by those unorthogonal vectors.
In general, for a Gaussian state represented by matrix $B$, we can obtain a canonical form for the representing matrix using the QR decomposition $B = Q \cdot R$, where $Q$ is a unitary matrix and $R$ is upper triangular.
Note that $Q$ and $B$ span the same linear space, so the Gaussian state can be expressed as $|Q\rangle$.
 
The supper operator $\mathcal{M}_{\Delta t}$ in Eq.~(\ref{apx:sse-trot}) corresponds to the Poisson jump process, where for each index $i$, we randomly decide whether a quantum jump process 
\begin{equation}
	|B\rangle \rightarrow \frac{L_i|B\rangle}{\Vert L_i|B\rangle \Vert}
\end{equation}
happens, with the probability 
\begin{equation}
	p_i = \langle B|L_i^\dagger L_i|B\rangle \gamma \Delta t
\end{equation}
The $L_m$'s we choose in the main text have the form 
\begin{equation}
	L_m = e^{ih} d^\dagger d,\quad 
	d^\dagger = \sum_i a_{i} c_i^\dagger,\quad
	h = \sum_{ij}h_{ij} c_i^\dagger c_j,
\end{equation}
where $d^\dagger$ is a quasimode, and $h$ is a fermion bilinear.
The following shows that the Gaussian form is preserved by such jump operator $L_m$.
First, the probability of the jump process is
\begin{equation*}
	\langle B|L_m^\dagger L_m|B\rangle
	= \langle B|d^\dagger d|B\rangle
	= \Vert d|B\rangle \Vert^2.
\end{equation*}
The action of annihilation operator $d$ on $|B\rangle$ is
\begin{equation}\label{eq:apx-action_of_d}
	d|B\rangle 
	= \sum_k a_k^* c_k \prod_{j} \sum_i c_i^\dagger B_{ij} |0\rangle 
	= \sum_j \langle a|B_j\rangle \bigotimes_{l\ne j}|B_l\rangle,
\end{equation}
so we can obtain the probability 
\begin{equation}
	p_m = \sum_j |\langle a|B_j\rangle|^2 \gamma \Delta t.
\end{equation}
Besides, we can utilize the gauge freedom to choose the basis such that $\langle a| B'_{j}\rangle = 0$ for $j>1$.
The matrix $B'$ always exists since we can always find a column $j$ that $\langle a| B_{j}\rangle \ne 0$ (otherwise, the probability of the jump is zero).
We then move the column to the first and define the column as
\begin{equation}
	|B'_{j}\rangle = |B_{j}\rangle - \frac{\langle a|B_{j}\rangle}{\langle a|B_{1}\rangle} |B_{1}\rangle, \quad j>1.
\end{equation}
Note that such column transformation does not alter the linear space $B$ spans, while the orthogonality and the normalization might be affected and should be renormalized afterward.
Eq.~(\ref{eq:apx-action_of_d}) then simplified to:
\begin{equation}
	d|B\rangle = \bigotimes_{j> 1} |B'_{j}\rangle.
\end{equation}
The result of the quantum jump is 
\begin{equation}
	L_m|B\rangle = |e^{ih}a\rangle \bigotimes_{j> 1} |e^{ih} B'_{j}\rangle.
\end{equation}
The representation of the outcome state is also not orthogonal.
An additional QR decomposition is needed to convert it to canonical form.

\section{Numerical simulations of the monitored free fermion dynamics}

\subsection{Projective monitoring}
\label{apx:proj-monitor}

\begin{figure*}
	\centering
	\includegraphics[width=\linewidth]{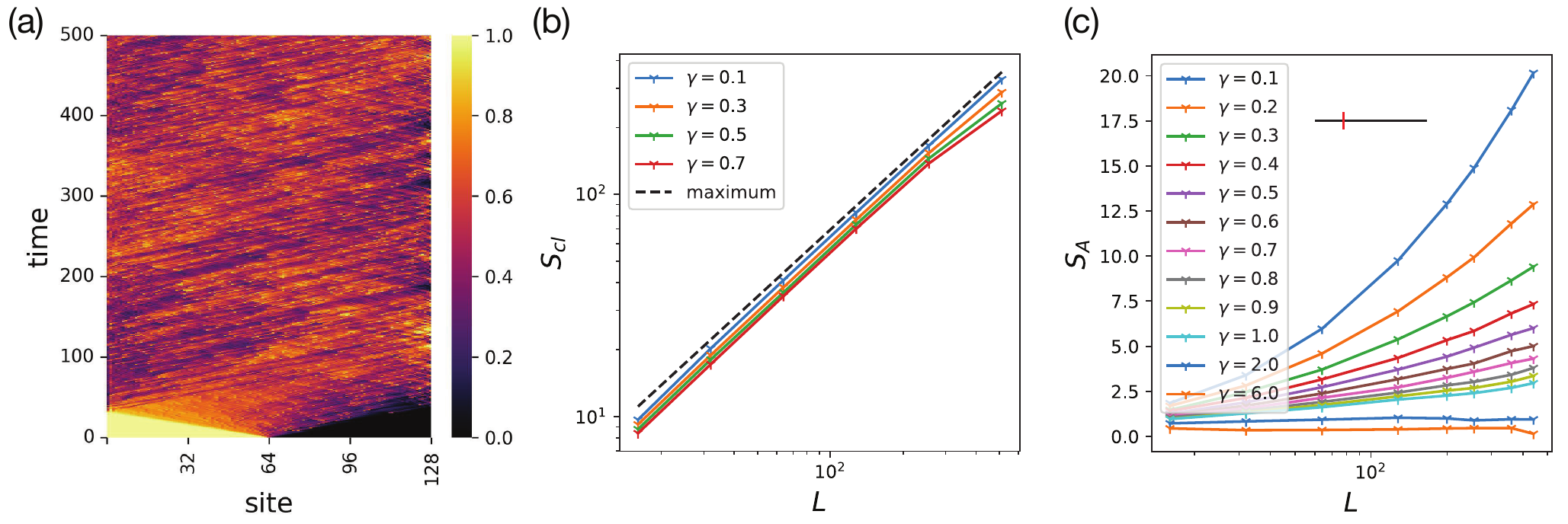}
	\caption{Numerical simulations of projective monitored fermion chains. (a) Density evolution under projective monitoring with $\gamma=0.1$ for a single trajectory. (b) Scaling of steady-state classical entropies. 
	(c) Trajectory average of entanglement entropy for different measurement rates. In particular, the entanglement entropy grows with system size in the small-$\gamma$ regime, while the entanglement entropy saturates to a constant value for large $\gamma$.}
\label{fig:C4}
\end{figure*}

We have demonstrated that a general Lindblad equation,
\begin{equation}
	\frac{d}{dt}\rho = -i\left[H,\rho\right] + \gamma \sum_i L_i \rho L_i^\dagger - \frac{1}{2} \left\{L_i^\dagger L_i, \rho\right\},
\end{equation}
where $L_i$ are projectors,
leads to a unique and homogeneous steady state (under open boundary conditions). 
In this section, we present numerical evidence supporting the homogeneity of late-time states for typical stochastic trajectories, described by:
\begin{equation}
\begin{aligned}
	d|\psi\rangle =&\ \left[-i H-\frac{\gamma}{2}\sum_m (L_m^\dagger L_m -\langle L_m^\dagger L_m\rangle)\right] |\psi\rangle dt \\
	&\ + \sum_m \left(\frac{L_m}{\sqrt{\langle L_m^\dagger L_m\rangle}}-1\right)|\psi\rangle dW_m.
\end{aligned}
\end{equation}
That is, the spatial homogeneity is not only at the density matrix level but also at the trajectory level.
This suggests that the skin effect, which suppresses entanglement entropy, does not arise within the projective monitoring system. 
Therefore, introducing conditional feedback is essential to induce a dynamical skin effect.

In the following, we consider the model studied in the main text, of which the unitary evolution is generated by the Hamiltonian:
\begin{equation}
	H = \sum_i (c^\dagger_i c_{i+1}+c^\dagger_{i+1}c_i)
\end{equation}
and the monitoring is described by
\begin{equation}
	L_i = \frac12 (c^\dagger_i-ic^\dagger_{i+1})(c_i+ic_{i+1}).
\end{equation}

First, Fig.~\ref{fig:C4}(a) displays the particle density evolution of a typical trajectory. 
At late times, the system reaches a homogeneous state with only minor fluctuations.
To further analyze the homogeneity of trajectories' densities, in Fig.~\ref{fig:C4}(b), we plot the trajectory-averaged classical entropy:
\begin{equation}
	S_\text{cl} = -\overline{\sum_i\left[ \langle n_i\rangle \log \langle n_i\rangle +(1-\langle n_i\rangle)\log(1-\langle n_i\rangle\right]}.
\end{equation} 
We observe a linear growth with system size, indicating a highly homogeneous density distribution. 
Moreover, when $\gamma$ is small, the growth of the classical entropy approaches the saturation value
\begin{equation}
	S_\text{max} = L \log 2,
\end{equation}
which means that most trajectory has nearly homogeneous density distribution.

In contrast, Fig.~\ref{fig:C4}(c) shows the entanglement entropy, which is considerably smaller than the classical entropy and exhibits an entanglement transition from log law to area law.  
These results suggest that the projective monitoring system under open boundary conditions may undergo a measurement-induced entanglement phase transition, similar to the periodic boundary condition.

\subsection{Generalized monitoring}
\label{numerics-gen}

\begin{figure*}
	\centering
	\includegraphics[width=\linewidth]{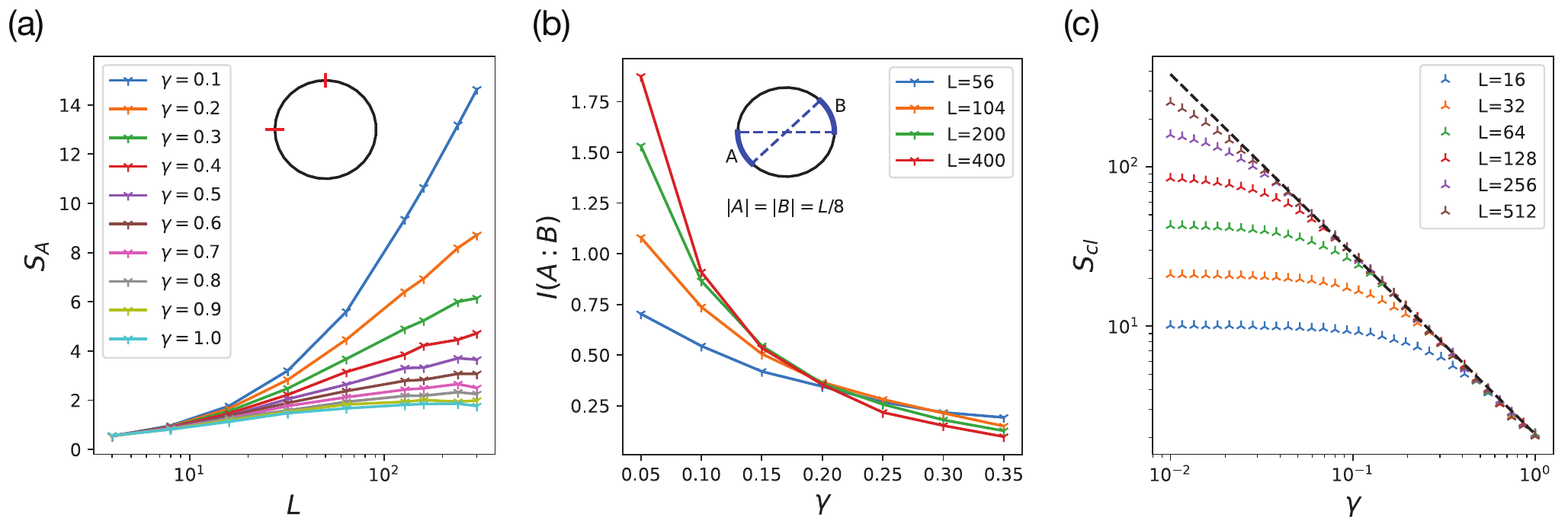}
	\caption{Numerical simulations of generalized monitored fermion chains. 	(a) Trajectory average of entanglement entropy for different measurement rates. (b) Trajectory average of mutual information $I(A:B)$ between two $L/8$ antipodal segments $A$ and $B$. The crossing of $I(A:B)$ indicates the critical measurement rate at $\gamma_c \approx 0.2$. (c) Trajectory average of classical entropy for systems in different sizes.}
\label{fig:C5}
\end{figure*}

Fig.~\ref{fig:C5}(a) illustrates the trajectory averaging of entanglement entropy. The entanglement entropy exhibits a transition from log law to area law.
We can pinpoint the critical measurement rate $\gamma_c$ by doing the scaling analysis of the mutual information between two antipodal regions (as displayed in Fig.~\ref{fig:C5}(b)).

In the main text, we have demonstrated trajectory averaging of classical entropy $\overline{S_\text{cl}[\{\langle n_i \rangle\}]}$ provides an upper bound on entanglement entropy $S_A$ and is itself bounded by classical entropy of averaged particle number:
\begin{equation}
	\overline{S_\text{cl}[\{\langle n_i \rangle\}]} \le S_\text{cl}[\{\overline{\langle n_i \rangle}\}]
\end{equation}
In this section, we instead calculate $\overline{S_\text{cl}[\{\langle n_i \rangle\}]}$ which imposes a tighter bound on the entanglement entropy.
As displayed in Fig.~\ref{fig:C5}(c), $\overline{S_\text{cl}[\{\langle n_i \rangle\}]}$ behaves similarly as $S_\text{cl}[\{\overline{\langle n_i \rangle}\}]$ but with slightly different slop, indicating an asymptotic form:
\begin{equation}
	\overline{S_\text{cl}[\{\langle n_i \rangle\}]} \simeq \frac{2.1}{\gamma^\nu}
\end{equation}
where the scaling factor $\nu \approx 1.13$ instead of $1$ for $S_\text{cl}[\{\overline{\langle n_i \rangle}\}]$.

\subsection{Generalized monitoring with different conditional feedback}
\label{different-theta}

\begin{figure*}
	\centering
	\includegraphics[width=\linewidth]{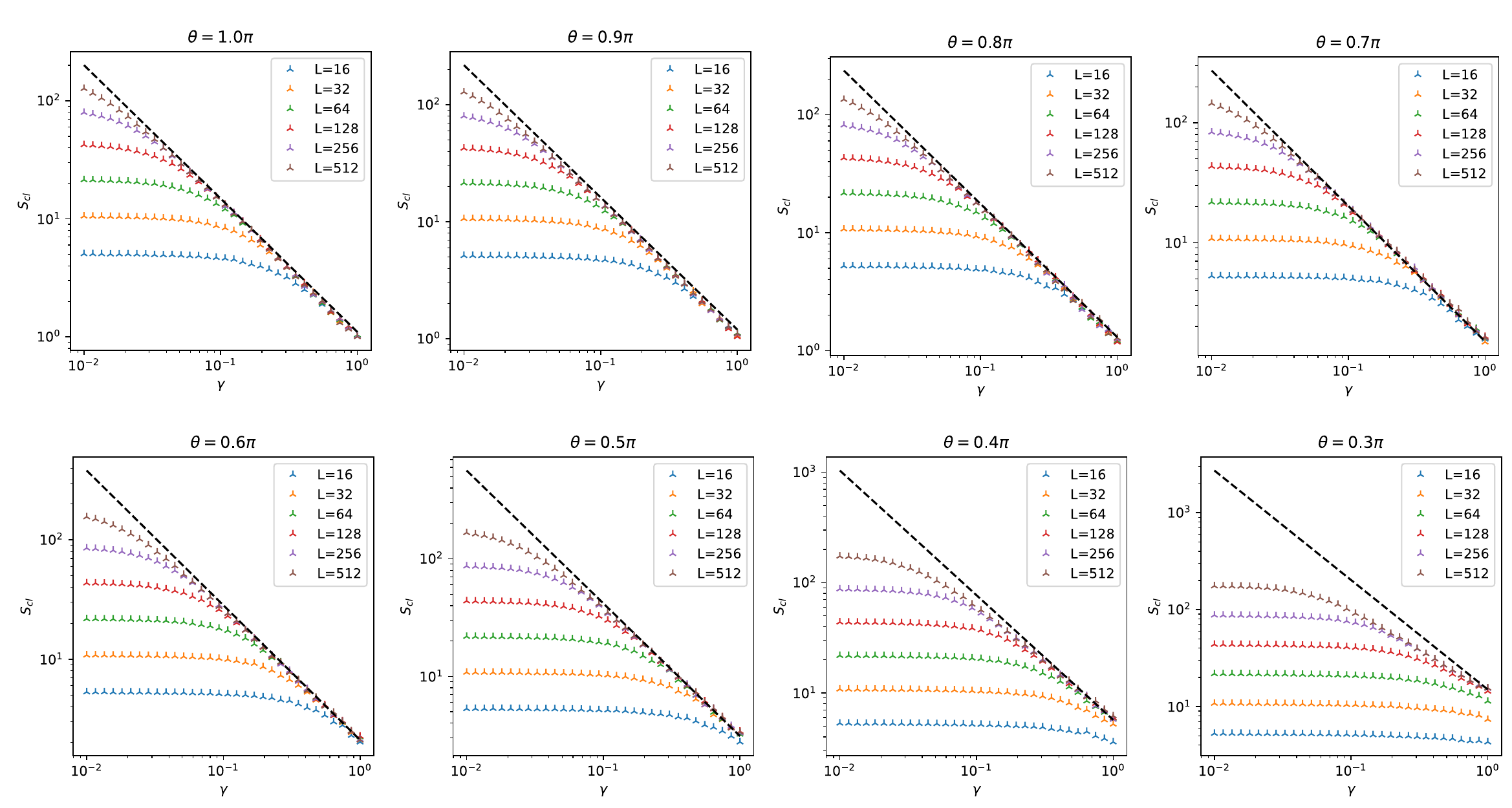}
	\caption{Steady-state classical entropies $\overline{S_\text{cl}[\langle n_i \rangle]}$ for feedback $U_i = \exp(i\theta n_{i+1})$, where we chose $\theta$ from $\pi$ to $0.3\pi$. the data points all approach the asymptotic line $\nu \log \gamma + \log S_\mathrm{cl} = c'(\theta)$ where $\nu\approx 1.13$ in the thermodynamic limit.}
	\label{Fig5}
\end{figure*}

In this appendix, we provide more numerical results on the monitored free fermion system with Hamiltonian
\begin{equation}
	H = \sum_i(c_i^\dagger c_{i+1} + c_{i+1}^\dagger c_i)
\end{equation}
and the generalized monitoring characterized by the projective monitoring 
\begin{equation}
	P_i = \frac{1}{2} (c_i^\dagger -i c_{i+1}^\dagger) (c_i+i c_{i+1})
\end{equation}
followed by the conditional $\theta$-phase rotation on site ($n+1$):
\begin{equation}
	U_i = \exp(i\theta n_{i+1}).
\end{equation}
In the main text, we show the extreme case where $\theta = \pi$ and argue that the feedback-induced skin effect persists for other choices of $\theta$ and therefore it is not a fine-tuned property of generalized monitoring.

Here we demonstrate the statement by numerical simulation of the trajectory-averaged classical entropy:
\begin{equation}
	S^\theta_\text{cl}(\gamma, L) \equiv \overline{S_\text{cl}[\{\langle n_i \rangle\}]}.
\end{equation}
Note that the quantity we show here is different from the main text, where we compute the classical entropy of the trajectory-averaged density. 
Here, instead, we calculate the classical entropy for different trajectories and then take the average over the trajectory ensemble.
This switch will not change the result qualitatively, while quantitatively this will lead to a smaller value and therefore easier for us to see the saturation in small $\theta$'s.
The quantity, however, has stronger fluctuation so it is hard to observe the scale invariance as in the main text.
Nevertheless, as we show in Fig.~\ref{Fig5}, while the $\theta = \pi$ case corresponds to an optimal angle where the skin effect is strongest, for other $\theta$ the asymptotic scaling
\begin{equation}\label{eq:apx-scaling-form}
	S^\theta_\mathrm{cl}(\gamma,L\rightarrow \infty) = \frac{c_\theta}{\gamma^{\nu}},
\end{equation}
persists, where $\nu \approx 1.13$, while the constant $c_\theta$ increases as $\theta$ decreases:
\begin{equation*}
\begin{tabular}{ccccccccc}
	\hline \hline
	$\theta/\pi$ & $1$ & $0.9$ & $0.8$ & $0.7$ & $0.6$ & $0.5$ & $0.4$ & $0.3$ \\ 
	\hline
	$\log c_\theta$ & $0.75$ & $0.85$ & $0.95$ & $1.1$ & $1.4$ & $1.9$ & $2.6$ & $3.4$ \\
	\hline \hline
\end{tabular}
\end{equation*}
As discussed in the main text, the constant $c_\theta$ determines the spatial extents of the domain wall.
The finite data point we obtained implies that $c_\theta$ increases exponentially fast when $\theta \rightarrow 0$.
In the $\theta=0$ case, $c_\theta$ diverges and there is no skin effect.

The large-size domain walls impose significant overhead for numerical simulations.
For small $\theta$ ($\le 0.3\pi$), accessing the thermodynamic limit becomes practically hard.

Based on the numerical results, we speculate the scaling form (\ref{eq:apx-scaling-form}) holds for different $\theta$ and $\gamma$.
Therefore we expect the skin effect to arise in a large class of generalized monitored systems.

\end{appendix}

\bibliography{ref}

\end{document}